\documentclass[fleqn,usenatbib]{mnras}

\usepackage{newtxtext,newtxmath}

\usepackage[T1]{fontenc}
\DeclareRobustCommand{\VAN}[3]{#2}
\let\VANthebibliography\thebibliography
\def\thebibliography{\DeclareRobustCommand{\VAN}[3]{##3}\VANthebibliography}

\usepackage{graphicx}	% Including figure files
\usepackage{amsmath}	% Advanced maths commands

\usepackage{amssymb}	% Extra maths symbols
\usepackage{tablefootnote}
\graphicspath{{Images/}}

\newcommand\nai{\ion{Na}{i}}

\newcommand\ali{\ion{Al}{i}}

\newcommand{\teff}{$T_{\rm eff}$}
\newcommand{\logg}{$\log g$}

\title[Bright metal-poor stars from the HESP-GOMPA survey]{A chemodynamical analysis of bright metal-poor stars from the HESP-GOMPA survey -- Indications of a non-prevailing site for light r-process elements}

\author[Bandyopadhyay et al. 2023]{
Avrajit Bandyopadhyay,$^{1,2,3}$\thanks{E-mail: abandyopadhyay@ufl.edu}
Timothy C Beers$^{2,4}$,
Rana Ezzeddine$^{1,2}$,
Thirupathi Sivarani,$^{3}$
\newauthor
Prasanta K Nayak,$^{5}$
Jeewan C Pandey,$^{6}$
Pallavi Saraf,$^{3,7}$ and
Antony Susmitha$^{3}$
\\
$^{1}$ Department of Astronomy, University of Florida, Gainesville, FL 32601, United States of America\\
$^{2}$ Joint Institute for Nuclear Astrophysics-Center for the Evolution of the Elements (JINA-CEE), USA\\
$^{3}$ Indian Institute of Astrophysics, Koramangala 2nd Block, Bangalore, 560034, India\\
$^{4}$ Department of Physics and Astronomy, University of Notre Dame, Notre Dame, IN 46556, United States of America\\
$^{5}$Instituto de Astrofísica, Pontificia Universidad Católica de Chile, Av. Vicuña MacKenna 4860, 7820436, Santiago, Chile\\
$^{6}$Aryabhatta Research Institute of Observational Sciences, Nainital 263001, India\\
$^{7}$Pondicherry University, R.V Nagar, Kalapet, 605014, Puducherry, India\\
}

\date{Accepted 26 February 2024. Received 27 September 2023; in original form 27 September 2023}

\pubyear{2024}

\begin{document}
\label{firstpage}
\pagerange{\pageref{firstpage}--\pageref{lastpage}}
\maketitle

\begin{abstract}

We present a comprehensive analysis of the detailed chemical abundances for a sample of 11 metal-poor, very metal-poor and extremely metal-poor stars ([Fe/H] = $-1.65$ to [Fe/H] = $-3.0$) as part of the HESP-GOMPA (Galactic survey Of Metal Poor stArs) survey. The abundance determinations encompass a range of elements, including C, Na, Mg, Al, Si, Ca, Sc, Ti, Cr, Mn, Fe, Co, Ni, Cu, Zn, Sr, and Ba, with a subset of the brighter objects allowing for the measurement of additional key elements. Notably, the abundance analysis of a relatively bright highly $r$-process-enhanced ($r$-II) star (SDSS J0019+3141) exhibits a predominantly main $r$-process signature and variations in the lighter $r$-process elements.  Moreover, successful measurements of thorium in this star facilitate stellar age determinations. We find a consistent odd-even nucleosynthesis pattern in these stars, aligning with expectations for their respective metallicity levels, thus implicating Type II supernovae as potential progenitors. From the interplay between the light and heavy $r$-process elements, we infer a diminishing relative production of light $r$-process elements with increasing Type II supernova contributions, challenging the notion that Type II supernovae are the primary source of these light $r$-process elements in the early Milky Way. A chemodynamical analysis based on Gaia astrometric data and our derived abundances indicates that all but one of our program stars are likely to be of accreted origin. Additionally, our examination of $\alpha$-poor stars underscores the occurrence of an early accretion event from a satellite on a prograde orbit, similar to that of the Galactic disc.
% Employing a %chemodynamical analysis based on Gaia data and the derived abundances, we establish that all but one star in %this study is likely to be of accreted origin.  Finally, we construct orbital trajectories spanning 12 Gyr into the past, providing valuable %insights into the historical dynamics of these stars and their contributions to the Galactic stellar populations.
% Tim -- NOTE THAT I SHORTENED THE ABSTRACT BY CUTTING OUT THE LAST FEW SENTENCES, EMPHASIZING OUR OTHER RESULTS.
\end{abstract}


%%%%%%%%%%%%%%%%%%%%%%%%%%%%%%%%%%%%%%%%%%%%%%%%%%

%%%%%%%%%%%%%%%%% BODY OF PAPER %%%%%%%%%%%%%%%%%%

\section{Introduction}
\label{sec:introduction}

Extremely metal-poor (EMP; [Fe/H] $\le -3.0$) stars are believed to be the immediate descendants of the first stars in the Galactic halo \citep{bromm2004,naoz,yoshida,Lazar2022,klessen2023,ji2024}. The evolution and subsequent explosions of these first stars, which were also massive, played a pivotal role in the genesis of all elements heavier than Li in the early Universe. The very metal-poor (VMP; [Fe/H] $\le -2.0$) stars also carry important imprints of the early stellar populations and are likely to be formed in the first epochs of the Universe \citep{cayrel2004,beers2005,bonifacio2007,bromm2009,cooke2014,frebelandnorris,frebelrev18}. 

The primordial supernovae from these early stars had profound effects on the dynamics and chemical evolution of the surrounding interstellar medium \citep{denissenkov96,sbordone2010,chiaki2017}. Subsequent generations of stars and their host galaxies are expected to bear the signatures of nucleosynthesis events originating from these Population III stars, e.g., as explored by \citet{nakamura1999}, \citet{frebelrev15}, \citet{sharma2018}, and \citet{Koutsouridou2023}. The wide diversity in the chemical composition of metal-poor stars has led to predictions and theoretical modelling regarding the possible nature and contributions from the progenitor populations \citep{hegerandwoosley2002,nomoto2013,Rossi2023}. The production sites of the different elements are largely decoupled from each other, and it is important to disentangle the contributions from the different types of progenitors 
(e.g., \citealt{gratton2004,barklem2005,matsuno2017,siegel2019}). 

New discoveries of metal-poor stars are not only crucial but also critical to quantify the elemental-abundance yields and carry out statistical studies to understand the astrophysical processes and environments in which they operate. 
The study of VMP and EMP stars has greatly benefited from comprehensive spectroscopic surveys conducted in the past to identify them in significant numbers. Notable examples of these early efforts include the HK survey of Beers and collaborators \citep{beers85,beers92} and the Hamburg/ESO Survey of Christlieb and colleagues \citep{christlieb2003}.

Over the past decade, $r$-process-enhanced (RPE) stars have emerged as a focal point of investigation among the diverse sub-classes of metal-poor stars.  Nevertheless, a profound understanding of the sources and astrophysical sites of $r$-process nucleosynthesis has remained elusive \citep{Schatz2022}. The synthesis of $r$-process nuclides requires environments capable of generating a substantial neutron flux within a specific entropy range, as described by, e.g.,  
\citet{argast2004}, \citet{hoto2015}, \citet{thielemann2017}, and \citet{Arcones2023}. Candidate astrophysical sites for $r$-process element production include core-collapse supernovae, magneto-rotationally driven supernovae, neutron star mergers, mergers involving neutron stars and black holes, and collapsars \citep{lattimer74,honda2006,roederer2010,frebelrev18,siegel2019}. 

The global effort of the 
$R$-Process Alliance (RPA; e.g., \citealt{hansen18rpa,sakari18rpa,rpa3,Holmbeck2020}) has been instrumental in increasing the number of known RPE stars and in providing a better understanding of the origin of $r$-process in stars. \citet{Shank2023} provides a recent compilation of the RPE stars reported in the literature and demonstrates, from an analysis of chemodynamically tagged groups (CDTGs) of RPE stars, that stars in these groups have likely experienced common chemical-evolution histories, presumably in their parent satellite galaxies or globular clusters, prior to being disrupted into the Milky Way's halo (see also \citealt{Gudin2021}).
Despite this progress, numerous open questions remain, including the apparent universality of the main $r$-process, obtaining a better understanding of the limited $r$-process, the origin of the so-called actinide-boost stars, and the nature and frequency of the progenitors, sites, and environments of $r$-process element production.

Along with the $r$-process elements, the abundances of other elements from carbon to zinc measured in metal-poor stars are crucial to derive and understand the nature of early supernovae. These elements are produced within the stars via stellar burning and during their explosions as supernovae of different mass and chemical composition (e.g., \citealt{truran2002,travaglio2004,tominaga2014,
jinmiyoon,Koutsouridou2023}). The unique abundance patterns of these elements help to understand the mixing of the ISM and the different astrophysical processes in operation at the time of a given star's formation. Such measurements in main-sequence stars take on particular importance, as they are the least affected by internal mixing and retain the original composition in the outer layers \citep{spite2013,mardini}. 

In addition to the chemistry of metal-poor stars, the 
Gaia mission \citep{gaia16} has provided the opportunity to understand the kinematics by studying the orbits and accretion history of these stars (e.g., \citealt{haywood2018, dimatteo2020, Limberg2021, Shank2022a, Shank2022b, Cabrera2023}). A more complete 
chemodynamical history of the stars can now be unravelled with Gaia DR3 \citep{gaiadr3} at much-improved precision and accuracy. This information allows for the determination of the present locations of these stars, and enables tracing of their dynamical histories, which is important to understand the formation and evolutionary history of the entire Galaxy.

In this paper, we report a chemical and chemodynamical analysis for 11 relatively bright MP, VMP, and EMP stars, including three main-sequence stars and a highly $r$-process-enhanced ($r$-II) giant star, identified from the MARVELS pre-survey effort as part of SDSS-III. The paper is organized as follows. In Section 2, we discuss the observational details. The spectroscopic analysis and kinematics are presented in Section 3. In Section 4, we discuss the abundances of all the detected elements in our program stars, while in Section 5, we discuss the implications and directions for future studies on metal-poor stars. We summarize and conclude in Section 6.

\section{Observations}
\label{sec:observation}

\subsection{Target Selection}

Our target stars were selected from the Multi-object APO Radial Velocity Exoplanets Large-area Survey (MARVELS) pre-survey data \citep{paegertmarvels}, which focused on spectroscopic radial-velocity measurements of multiple relatively bright objects ($V \leq$ 12.50) for efficient exoplanet discovery within the framework of SLoan Digital Sky Survey (SDSS) III \citep{eisenstein}. Drawing from our previously published work \citep{susmitha, bandyopadhyay,ban_gce,banli}, we employed synthetic spectral-fitting techniques to stars of the pre-survey data to identify likely metal-poor candidates. The most metal-poor targets based on the analysis were observed as a part of Hanle Echelle SPectrograph-Galactic survey Of Metal Poor stArs (HESP-GOMPA) survey.

\subsection{Data Reduction}

Spectroscopic observations of the candidate metal-poor stars were conducted between 2018 and 2021 using the Hanle Echelle Spectrograph (HESP) on the 2-m class Himalayan Chandra telescope (HCT) at the Indian Astronomical Observatory (IAO), yielding high-resolution spectra ($R \sim$ 30,000). The wavelength range for observations with HESP spans from 3500\,{\AA} to 10000\,{\AA} in a single exposure, without any discontinuities or gaps. Detailed information on the apparent magnitude, spectral signal-to-noise ratios (SNRs) at 5000 \AA, exposure times, dates of the observations, and radial velocities (RVs) can be found in Table 1.

The data processing was conducted with the IRAF echelle package \citep{tody1986,tody1993}. The HESP instrument features a dual-fibre mode, with one fibre focused on the target star and the other capable of receiving a calibration source for precise radial velocity measurements or capturing night-sky signals. In our application, the sky fibre served its purpose of efficiently subtracting the signature of the background sky. Wavelength calibration spectra were acquired from Th-Ar arc lamps following the completion of each observation. All spectral orders underwent normalization, RV correction, and subsequent merging to generate the final composite spectrum. A customized (publicly available) data-reduction pipeline for HESP was used.  This pipeline is designed to handle the complex distorted spectral orders resulting from HESP's observations of stellar spectra, which tend to be crowded together, particularly at the blue end. 

\subsection{Radial Velocity}

The HESP instrument maintains precise thermal control throughout the year which yields a sustained stability level of approximately 200 m/s, significantly mitigating systematic errors. To derive the RVs for each spectrum, a cross-correlation analysis was conducted employing a synthetic template spectrum tailored to each specific star. The derived RVs and the modified Julian dates (MJDs) for the observation epochs are listed in Table 1, along with the RVs obtained from Gaia. We note a significant deviation of the derived RV for the star SDSS J0216+3310 from the Gaia RV, with a difference of 52 km/s. It could be a possible binary, and multi-epoch observations are necessary to understand the nature of the binarity.  Smaller deviations, up to 10 km/s, are noted for SDSS J0019+3141 and SDSS 0210+3220.

\begin{table*}
	\centering
	\caption{Details and observations of the program stars.}
	\label{tab:object_table}
	\begin{tabular}{lccccrccrr} 
		\hline
		Star name &Object &RA & DEC & $g$ mag & SNR &Exposure &MJD & RV Gaia &RV HESP \\
                  &       &   &     &         &     & (sec)            &    &  (km/s) &  (km/s)\\
		\hline
SDSS~J001931.76+314144.1 &J0019+3141 &00:19:31.76 &+31:41:44.1 &9.72 &58 &7200 &57750 &$-$0.20  &-4.7\\
SDSS~J002205.84+321251.6 &J0022+3212 &00:22:05.84 &+32:12:51.6 &11.83 &21 &8100 &58009 &$-$64.16 &-64.2\\
SDSS J021053.91+322031.1 &J0210+3220 &02:10:53.91 &+32 20:31.1 &12.55 &34 &8100 &58068 &22.39 &31.0\\
SDSS~J021650.42+331026.7 &J0216+3310 &02:16:50.42 &+33:10:26.7 &11.31 &19 &8100 &58021 &30.38 &$-$22.1\\
SDSS J044703.20+542634.7 &J0447+5426 &04:47:03.20 &+54:26:34.7 &12.14 &33 &8100 &58097 &$-$292.90 &$-$293.8\\
SDSS J075305.22+490853.8 &J0753+4908 &07:53:05.22 &+49:08:53.8 &11.19 &22 &8100 &58133 &2.11 &0.57\\
SDSS~J135058.37+481917.0 &J1350+4819 &13:50:58.37 &+48 19:17.0 &12.55 &36 &8100 &57857 &$-$115.02 &$-$116.5\\
SDSS~J152147.09+364730.8 &J1521+3647 &15:21:47.09 &+36:47:30.8 &12.43 &23 &8100 &58133 &$-$52.40 &$-$54.7\\
SDSS~J193018.91+692636.1 &J1930+6926 &19:30:18.91 &+69:26:36.1 &12.49 &36 &8100 &58045 &$-$127.27 &$-$125.8\\
SDSS~J195344.22+422249.9 &J1953+4222 &19:53:44.22 &+42:22:49.9 &9.03 &85 &7200 &58004 &$...$ &$-$315.3\\
SDSS~J232030.34+174237.4 &J2320+1742 &23:20:30.34 &+17:42:37.4 &12.26 &37 &8100 &58068 &$-$311.43 &$-$311.4\\

		\hline
	\end{tabular}
\end{table*}

\section{Analysis}

\subsection{Stellar Parameters}

\label{sec:line_list_stellar_paramters}

We employ one-dimensional LTE stellar atmospheric models (ATLAS9; \citealt{castellikurucz}) and the spectral-synthesis tool TURBOSPECTRUM \citep{alvarezplez1998} to determine the elemental abundances in each spectrum. Absorption lines with equivalent widths (EWs) exceeding 130 m{\AA} and below 5 m{\AA} were not considered, as they are beyond the linear portion of the curve of growth,  and provide unreliable estimates of the corresponding abundances. For this analysis, we use Version 12 of the TURBOSPECTRUM code, incorporating the Kurucz linelist database and adopting hyperfine splitting data, as well as Solar isotopic ratios from \citet{asplund2009}.

To estimate the stellar atmospheric parameters, we followed an iterative approach as discussed in \citet{bandyopadhyay,ban_gce,banli}. Initial estimates for effective temperature were derived from photometric colors, specifically $V - K$, which are known to be least affected by the metallicity of the star. Following the first estimate from photometry, spectroscopic techniques were employed to deduce the finally adopted effective temperatures. A comprehensive grid of stellar models spanning various \teff, log $g$, and [Fe/H] values was prepared. We evaluated the abundances of clean Fe I and Fe II lines for each spectrum using EW analysis. Our fitting procedure ensured that the Fe I abundances remained consistent across excitation potential and matched the values obtained from Fe II lines. The wings of the  H-$\alpha$ line also exhibit sensitivity to subtle temperature fluctuations for metal-poor stars, as demonstrated in the upper panels of Figure 1. We used synthetic-template spectra of varying temperatures to fit the  H$\alpha$ wings, and the best fit was adopted. In this figure, the red lines indicate the best fit, while the blue and green lines indicate deviations for temperature differences of $\pm 150$\,K.

A similar approach was taken for the measurement of \logg, based on the Mg spectral features.  The \logg\ was determined through spectral fitting of the Mg I triplet wings in the 5172\,{\AA} to 5183\,{\AA} region, as the wings of these broad features are sensitive to minor variations in pressure, and thus serve as a useful indicator of surface gravity. In the bottom three panels of Figure 1, we showcase examples of fitting the Mg triplet wings. The red lines indicate the best fit, and the other colours indicate deviations from the adopted best-fit values by $\pm 0.50$ dex. Independent \logg\ estimates were carried out through (i) the ionization-equilibrium technique by demanding the same abundance values from the Fe I and Fe II lines, and (ii) using Gaia parallax values as described below. We calculated the surface gravity \logg\ using the relation:  log($g/g_{\odot}$) = log(M/M$_{\odot}$) + 4log(T$_{\rm eff}$/T$_{\rm eff\odot}$) + 0.4(M$_{bol}$ $-$ M$_{bol\odot}$). However, we have adopted the pressure broadened wings of the Mg I triplet as the finally chosen estimate for logg as they are very sensitive for minor variations of logg in case of cool stars (T < 6200 K) as discussed in \citet{Alexeeva2018} and \citet{Brewer2016}. All our stars are cooler, so we expect this to be the best indicator of logg. The logg obtained from Fe II lines and Gaia are consistent with the Mg I lines up to a deviation of 0.30 dex.

The $V$ magnitudes were sourced from SIMBAD, and parallaxes were extracted from Gaia when available. Evolutionary tracks were employed to estimate stellar masses, yielding values around 0.8 $M_{\odot}$ for these metal-poor stars. These parameters converged within a few iterations and remained consistent thereafter, with typical uncertainties of approximately 150\,K for \teff\ and 0.50 dex for $\log g$. 

The microturbulent velocity ($\xi$) for each star was determined iteratively by ensuring the absence of a correlation between the Fe I abundances and reduced equivalent widths. The final adopted atmospheric parameters for the stars are listed in Table 2.

\begin{figure*}
\centering
\includegraphics[width=1.8\columnwidth]{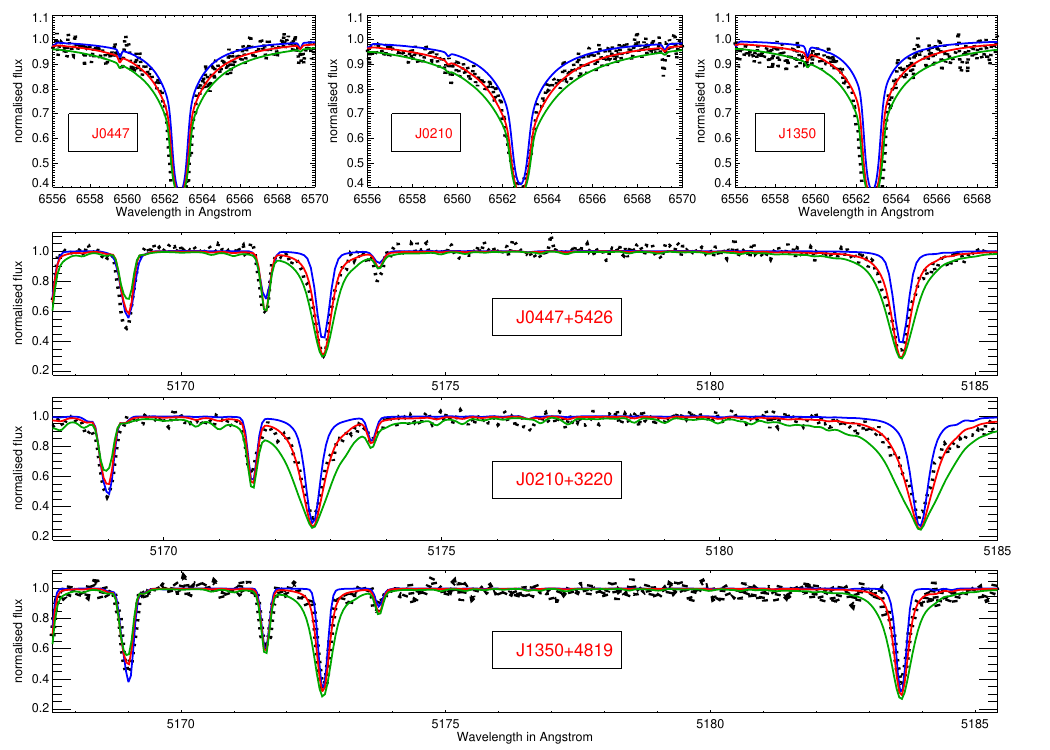}
\caption{An example for the fits to the H-$\alpha$ wings for determination of temperature and the Mg I triplet wings for determination of $\log g$. The best fit is shown in red; departures by $\pm 150$\,K and $\pm 0.50$ dex are shown in the coloured lines in the top and bottom panels, respectively.}
\end{figure*}

\begin{table}
	\centering
	\caption{Final adopted stellar parameters for our program stars.}
	\label{tab:Adopted_stellar_parameters}
	\begin{tabular}{lccccc}
		\hline
		Star name & $T_{\rm eff}$ & $\log g$ &  $\xi$ & [Fe/H] \\
		 & (K) & (cgs)  & (km/s) & \\
		\hline
SDSS~J0019+3141 &4400 &1.00 &1.90 &$-$2.30\\
SDSS J0022+3212 &4600 &3.00 &1.95 &$-$1.60\\
SDSS J0210+3220 &5200 &4.00 &1.65 &$-$2.40\\
SDSS J0216+3310 &5400 &4.50 &1.20 &$-$1.70\\
SDSS J0447+5426 &5200 &3.70 &1.90 &$-$2.50\\
SDSS J0753+4908 &5100 &4.50 &2.10 &$-$2.05\\
SDSS~J1350+4819 &5100 &2.50 &1.20 &$-$2.90\\
SDSS J1521+3647 &5200 &3.50 &1.75 &$-$2.10\\
SDSS J1930+6926 &4600 &2.10 &1.50 &$-$3.00\\
SDSS~J1953+4222 &6000 &4.00 &1.75 &$-$2.25\\
SDSS J2320+1742 &4800 &2.20 &2.20 &$-$2.30\\

        \hline
	\end{tabular}
\end{table}

\subsection{Kinematics}

To estimate kinematic parameters for the observed stars, we require precision data on distances, proper motions, and RVs. Fortunately, Gaia \citep{gaia16} provides high-precision proper motions, which we adopted from Gaia-DR3 \citep{gaiadr3}. Distances were acquired from the 
\citet{bailerjones2021} catalogue, where the authors leveraged Bayesian methods to estimate distances based on Gaia-EDR3 parallaxes \citep{gaia_edr3}. Radial velocities were deduced from the observed spectra, as listed in Table 1. Our conversion into the Galactocentric system and the calculation of Galactocentric distances (X, Y, Z) and velocities ($V_x$,$V_y$, $V_z$) were obtained using the Astropy module as discussed in \citet{banli}. Here, the Galactic plane corresponds to the X-Y plane. This conversion considered the Sun's Galactocentric distance as 8.2 kpc \citep{dehnenbinney98,mcmillan2017} and the velocity of the Local Standard of Rest $V_{LSR}$ = 233.1 km/s. We take the Sun's distance from the Galactic plane as 0 kpc and the Solar velocity components as (12.9, 245.6, 7.78) km/s, as discussed in Appendix B of \citet{meingast21}.

\begin{table*}
\tabcolsep6.0pt $ $
\begin{center}$ $
\caption{Kinematics for the program stars.}
\begin{tabular}{crrrrrrrcccccccr}
\hline\hline
Object  &Distance  &X  &Y  &Z  & $V_x$  & $V_y$  & $V_z$ &$L_{\perp}$  & $L_z$  & $V_{\phi}$  &Energy &Eccentricity \\
        &(kpc) &(kpc) &(kpc) & (kpc) & (km/s) & (km/s) & (km/s) & (10$^2$ kpc km/s) & (10$^2$ kpc km/s) & (km/s) & (10$^4$ km$^2$s$^{-2}$) \\   
        \hline
            
J0019+3141 &1.6 &$-$8.7 &1.3 &$-$8.5 &$-$235.2 &79.7     &$-$67.5 &3.93 &$-$3.98    &$-$44.0  &$-$12.06 &0.88\\
J0022+3212 &1.5 &$-$8.4 &1.1 &$-$7.7 &10.0     &158.5    &2.9     &1.27 &$-$14.23   &$-$158.4 &$-$14.20 &0.33\\
J0210+3220 &0.4 &$-$8.3 &2.3 &$-$1.9 &$-$27.5  &226.1    &$-$2.7  &0.46 &$-$19.39   &$-$225.3 &$-$13.13 &0.11\\
J0216+3310 &0.1 &$-$8.2 &9.4 &$-$0.1 &$-$0.7   &238.6    &$-$47.1 &3.88 &$-$20.13   &$-$238.6 &$-$12.87 &0.06\\
J0447+5426 &0.6 &$-$8.6 &2.8 &0.0    &246.5    &72.5     &$-$67.0 &5.64 &$-$7.15    &$-$80.6  &$-$12.02 &0.82\\
J0753+4908 &0.2 &$-$8.2 &3.5 &0.1    &40.0     &240.6    &59.8    &5.01 &$-$20.47   &$-$240.7 &$-$12.63 &0.14\\
J1350+4819 &1.9 &$-$8.2 &7.9 &1.8    &$-$229.8 &11.4     &$-$30.0 &6.62 &0.92       &10.90    &$-$12.78 &0.88\\
J1521+3647 &2.4 &$-$7.4 &1.1 &2.0    &2.3      &152.3    &1.0     &3.13 &$-$11.60   &$-$150.8 &$-$14.78 &0.33\\
J1930+6926 &2.9 &$-$8.6 &2.6 &1.1    &50.5     &115.0    &3.1     &1.44 &$-$11.55   &$-$124.8 &$-$14.47 &0.47\\
J1953+4222 &0.1 &$-$8.0 &1.2 &0.0    &$-$118.2 &$-$34.2  &34.5    &2.77 &2.99       &36.1     &$-$15.17 &0.86\\
J2320+1742 &3.7 &$-$8.3 &2.8 &$-$2.3 &89.0     &$-$192.3 &$-$34.6 &7.48 &13.86      &153.3    &$-$12.78 &0.48\\

\hline
\end{tabular}
\end{center}
\end{table*}

%IN THE KINEMATICS TABLE, GIVE UNITS FOR ENERGY AND TWO DECIMAL ACCURACY FOR ECCENTRICITY. AND USE $-$ FOR NEGATIVE QUANTITIES.  ANGULAR MOMENTA AND ENERGY CAN INCLUDE A FEW DECIMALS ACCURACY RATHER THAN 1

We computed the orbital characteristics of the program stars using methods following 
\citet{pinto2021} and \citet{banli}, assuming a four-component potential model for the Milky Way. This model comprises a bulge and nucleus following the \citet{Hernquist} profile, a disk adhering to the Miyamoto-Nagai \citep{MiyamotoNagai} profile, and a halo represented by an NFW \citep{NFW} profile. For consistency with prior research, we have held the parameters of the disk and bulge constant, aligning them with the values established in \citet{Bovy}.  In our computations, the Galactocentric distance was represented by $r$. The parameter $V_r$ indicated the velocity component along $R$, while $V_z$ represented the vertical velocity component of the stars. The angular momentum's $z$ component is denoted as $L_z$, and the perpendicular component of the angular momentum is $L_{\perp}$. The azimuthal velocity, $V_{\phi}$, is defined as $L_z/R$. All computed velocities and angular momenta for the program stars are listed in Table 3, with velocities in km/s, angular momenta in units of 10$^2$ kpc km/s, and specific energy in units of 10$^5$ km$^2$s$^{-2}$.

\begin{figure}
\centering
\includegraphics[width=1.0\columnwidth]{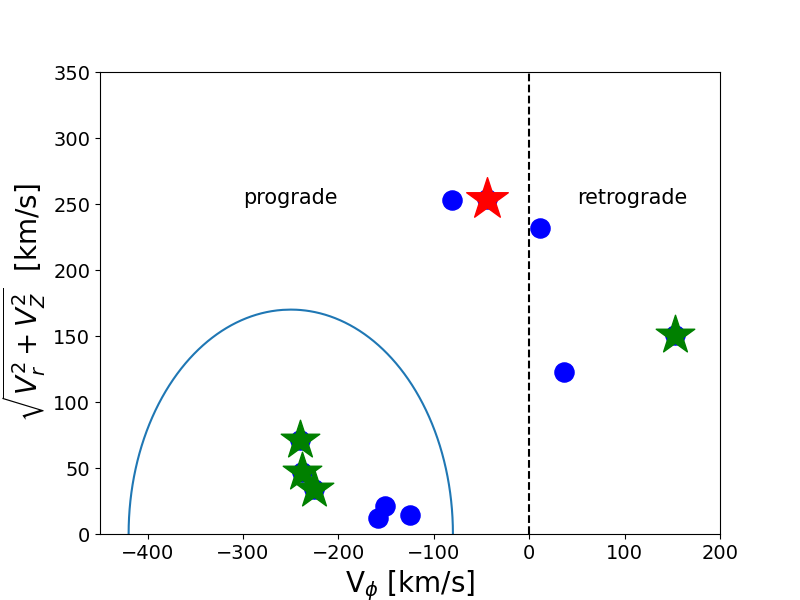}
\caption{Toomre Diagram showing the orbital characteristics of the program stars. The vertical dashed line separates prograde motions from retrograde motions, while the blue semicircle in the bottom left indicates stars with disc-like orbits. The $r$-II star is shown in red, while the $\alpha$-poor stars are shown in green. The other stars are shown in blue.}
\label{Vper_Vphi}
\end{figure}

Following the conventions of \cite{dimatteo2020}, where a clockwise rotation of the disc was assumed, a negative value of $V_{\phi}$ indicates prograde motion, while a positive value indicates retrograde motion. As depicted in the Toomre Diagram shown in \autoref{Vper_Vphi}, the majority of our program stars exhibit prograde motions, with the vertical dashed line distinguishing between prograde and retrograde motions. The blue semicircle outlines the anticipated position of disc stars in this plane. Six stars are likely to be members of the disc system and fall within the semicircular patch shown in the lower left. Three $\alpha$-poor stars are also seen to have disc-like orbital characteristics with prograde motions. The star marked in red is the $r$-II star SDSS J0019+3141, a halo member with a likely prograde motion (see section 5.3).

\section{Elemental Abundances}\label{sec:abund}

We determine the abundances of C, Na, Mg, Al, Si, Ca, Ti, Sc, Cr, Mn, Co, Ni, Cu, Zn, Sr, and Ba in most of the stars in this study. Some additional elements like Eu, La and other r-process elements could be derived for two of the stars. The abundances for the elements measured in each star are tabulated in tables 5-15. The columns indicate the name of the element, species, number of lines, absolute abundance of the measured element A(X), solar abundance, abundance with respect to hydrogen ([X/H]), abundance with respect to Fe ([X/Fe]) and the uncertainty associated with the measurement. The values for the solar abundances have been obtained from \citet{asplund2009}  The abundances of the individual elements are discussed below.

\subsection{Carbon}

Carbon (C) holds significant importance in the context of metal-poor stars due to its formation through various processes within massive stars and early supernovae \citep{bonifacio2015,jinmiyoon,placcocemp2014,Rossi2023}, with significant implications for star formation and chemical evolution in the early Galaxy. Additionally, carbon is produced in low- to intermediate-mass AGB stars \citep{Cristallo16,sus21} , and plays a crucial role in distinguishing different stellar populations in the metal-poor regime. To determine the carbon abundances in our program stars, we employ an iterative approach involving the fitting of the molecular CH $G$-band around 4315\,{\AA} using spectrum synthesis, as described by \citet{masseron2014}. An example of carbon synthesis is presented in Figure 3 for three program stars. In the figure, the red line represents the best-fit carbon abundances, while deviations of $\pm 0.50$ dex are indicated with blue and green lines. The range of carbon abundances among these stars spans from [C/Fe] = $-$0.88 to [C/Fe] = $+$0.67. As the majority of these stars are in the red giant phase, we apply corrections to the measured carbon abundances to account for evolutionary effects, as provided by \citet{placco2014}. Note that the corrections for the main sequence or main-sequence turnoff stars are minimal. The carbon abundances, along with other elements, are shown in Figure 4.  

%THE TABLES LISTING THE ABUNDANCES FOR CARBON SHOULD INCLUDE BOTH THE MEASURED AND CORRECTED CARBON ABUNDANCES.

%Since the corrections for all the stars except J0019 are very low (less than 0.05 dex), I havent listed the corrections separately.

\begin{figure*}
\centering
\includegraphics[width=1.5\columnwidth]{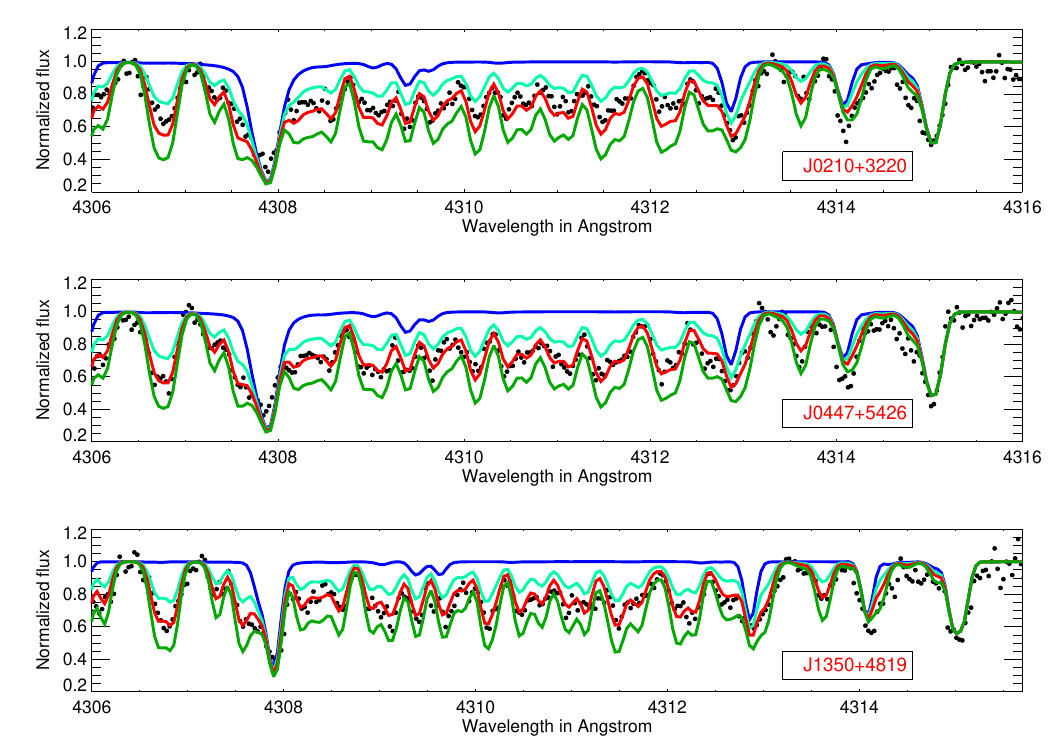}
\caption{An example for the determination of the carbon abundance from the molecular CH $G$-band. The red line indicates the best fit to the observed points shown in black, while the light blue and green lines correspond to deviations of 0.3 dex from the best fit. The blue line denotes the synthetic spectrum for complete absence of carbon.}
\end{figure*}

\subsection{Odd-Z Elements}
\subsubsection{Sodium}

The origin of sodium (Na) is primarily from massive stars, where it is synthesized through processes such as hydrostatic carbon burning and hydrogen burning in the Ne-Na cycle \citep{christal2015}. In our analysis, we determined Na abundances by measuring the \nai\ doublet D1 and D2 lines located at 5889 and 5895\,{\AA}, respectively. We considered non-LTE corrections for Na, as computed by \cite{andrievskyna}, which typically amounted to about 0.10 dex, but occasionally extended up to 0.20 dex, within the given metallicity range. The Na abundances for the program stars in this study are corrected for NLTE effects by subtracting an average of 0.15 dex from the measured abundances and the corrected values are listed in the abundance tables.

\subsubsection{Aluminium}

Aluminium (Al) is primarily synthesized in evolved massive stars through processes involving carbon and neon burning \citep{nomoto2013}. We estimated Al abundances by analyzing the resonance \ali\ line at 3961\,{\AA}. Non-LTE corrections for Al were established by \cite{andrievskyal} and \cite{nordlander2017}, with potential deviations as high as 0.70 dex within the considered metallicity range. The corrected Al abundances for the program stars by adding 0.65 dex to the measured values are listed in the abundance tables. Notably, among the light elements, we observed significant scatter in the Al abundances, primarily attributed to increased errors stemming from inadequate SNR in the blue spectral region. As a result, a substantial fraction of the stars in our sample have undetectable Al. The Al abundances for the program stars for which they are measured are sub-solar, as expected for metal-poor stars; the distribution is shown in Figure 4.

\subsection{$\alpha$-Elements}

\begin{figure*}
\centering
\includegraphics[width=2.15\columnwidth]{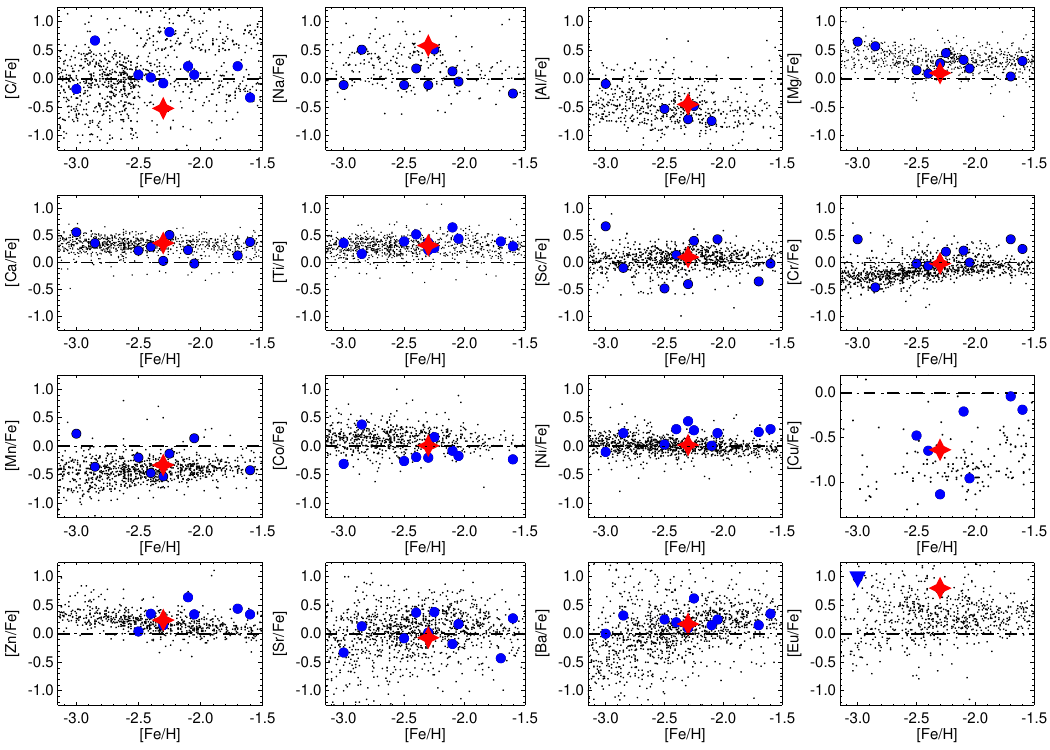}
\caption{The distribution of abundances for the detected elements in all program stars as a function of [Fe/H]. The blue-filled circles represent the program stars, while the red-filled star shows the abundance for SDSS J0019+3141. The abundances of other metal-poor stars in the literature, compiled from the SAGA database \citep{sudasaga}, are shown in black dots.  The dashed line indicates the Solar values.}
\end{figure*}

\subsubsection{Magnesium}

Magnesium (Mg) is produced during hydrostatic carbon burning in the shells of massive stars and prodigiously in Type Ia supernovae explosions \citep{Kobayashi2020}. In our investigation, we determined magnesium abundances by analyzing the EWs of specific non-blended Mg lines. Metal-poor stars exhibited a slightly elevated mean value of $\langle$[Mg/Fe]$\rangle = +0.35$, consistent with expectations for metal-poor halo stars, showing reduced scatter around this mean. Notably, a few of the program stars exhibit very low Mg abundances, including the $r$-II star SDSS J0019+3141, as seen in the trends in Figure 4. These $\alpha$-poor stars are interesting candidates for follow-up studies. They are likely to be accreted; their kinematics are discussed in sub-section 5.3.

\subsubsection{Silicon}

Silicon (Si) is produced both during the pre-explosive and explosive phases of core-collapse supernovae, involving oxygen- and neon-burning processes \citep{hegerandwoosley2002}. In our study, SNR limitations in the spectra prevented us from utilizing the prominent Si transition line at 
3905\,{\AA}. Consequently, depending upon the weaker lines, we could only detect Si for two program stars, SDSS J0019+3141 and SDSS J0216+3310.

\subsubsection{Calcium}

Calcium (Ca) is another crucial $\alpha$-element indicator, and we successfully detected several Ca transitions across the spectra for all stars. The derived abundance ratios ranged between [Ca/Fe] =  0.0 and [Ca/Fe] = +0.60. The mean abundance ratio for the program stars, $\langle$[Ca/Fe]$\rangle = +0.34$,  matches with the expected halo-star enhancement of [$\alpha$/Fe] = $+$0.40.

\subsubsection{Titanium}

Significant amounts of titanium (Ti) are believed to be ejected during core-collapse supernovae and hypernovae explosions. We successfully identified clean features of both Ti I and Ti II for all the program stars in our study. The mean Ti abundance, $\langle$[Ti/Fe]$\rangle = +0.34$, closely matches the $\alpha$-enhancement ratios observed in halo stars. The agreement between Ti I and Ti II values remains within a reasonable range of up to 0.20 dex. 

\subsection{Fe-Peak Elements}
\subsubsection{Scandium}

Scandium (Sc) abundances were determined from multiple lines, with particular emphasis on the prominent transition at 4254\,{\AA}. However, the Sc abundances among the program stars exhibit significant scatter, suggesting contributions from supernovae originating with diverse masses in the precursor gas cloud. They do not exhibit any strong trends with metallicity.

\subsubsection{Chromium}

Chromium (Cr) is synthesized in the incomplete explosive combustion of silicon in Type II supernovae. Multiple Cr I lines, including the stronger ones at 4646\,{\AA} and 5206\,{\AA}, are detected within the spectra. However, these lines are also subject to significant NLTE corrections \citep{bergemanncescutti2010}. Cr II lines were also measured in select evolved stars, revealing a mean difference of 0.25 dex between the Cr I and Cr II lines in our sample, consistent with previous studies (e.g., \citealt{bonifacio2009}).

\subsubsection{Manganese}

Manganese (Mn) is another product of incomplete silicon burning in core-collapse supernovae \citep{nakamura1999}. The Mn abundances for most of our stars are determined using the resonance Mn I triplet at 4030\,{\AA}, complemented by an additional line at 4823\,{\AA}. We considered weaker features only when the SNR in the 4030\,{\AA} region was insufficient for reliable abundance measurements. However, it is important to note that these lines are susceptible to both 3D and NLTE corrections, which can range from 0.3 to 0.6 dex, as reported by \cite{bergemann2019}. No Mn II lines are detected in our spectra.

\subsubsection{Cobalt}

The \textsuperscript{59}Cu isotope, which subsequently decays into \textsuperscript{59}Co, originates from the complete silicon burning in core-collapse supernovae. This is the only stable isotope of Co that can be measured in metal-poor stars \citep{nakamura1999}. The Co abundances in our study are mainly derived from strong features at 3995\,{\AA} and 4121\,{\AA}, and they follow the expected trends for metal-poor stars.

\subsubsection{Nickel}

The [Ni/Fe] ratio primarily reflects the \textsuperscript{58}Ni/\textsuperscript{56}Ni ratio, with the dominant source of \textsuperscript{56}Ni from complete silicon-burning regions in core-collapse supernovae, influenced by the depth of the gravitational potential and neutrino-absorbing matter. Nickel (Ni) tends to track the Fe content and has been detected for all our program stars. The mean abundance ratio, $\langle$[Ni/Fe]$\rangle$ = +0.15. for our sample aligns well with the expected relation for the same production region of Ni and Fe. Notably, this ratio remains relatively stable as Ni and Fe are synthesized within the same region, making their ratios resistant to alteration \citep{Kobayashi2020}. 

\subsubsection{Copper}

Copper (Cu) is generated through the $s$-process in both massive and low-mass stars, with a significant contribution from explosive nucleosynthesis in Type II supernovae \citep{korotin}. Measuring Cu abundances in our program stars is challenging due to the low SNR; we could only detect Cu lines in 8 stars using the 5105.5\,{\AA} line. The abundance ratios for the metal-poor stars in our sample range between [Cu/Fe] = -0.5 and [Cu/Fe] = $+0.5$, consistent with previous findings.

\subsubsection{Zinc}

Zinc (Zn) originates in the deepest layers of core-collapse supernovae, with higher explosion energy leading to enhanced Zn production in hypernovae \citep{Kobayashi2020}. As a result, Zn serves as a crucial element for constraining progenitor supernovae, and is detected for the majority of our program stars. Our analysis relied on only two useful Zn lines at 4722\,{\AA} and 4810\,{\AA} to determine abundances. NLTE corrections for Zn are less than 0.1 dex \citep{sitnova}. 

\subsection{Neutron-capture Elements}

Astrophysical sites for the production of neutron-capture elements via $r$-process are still debated -- the best candidates being neutron star mergers, neutron star black hole mergers, and supernovae with jets. The relative abundances of the different neutron-capture elements are used to constrain the possible progenitors. We determine the abundances of a number of these elements, including 
strontium (Sr), yttrium (Y), zirconium (Zr), barium (Ba), cerium (Ce), neodymium (Nd), samarium (Sm), europium (Eu), dysprosium (Dy), and thorium (Th), using spectral-synthesis techniques. We carefully considered hyperfine transitions in our analysis, following the methodologies outlined in studies by \citet{mcwilliam1998} and \citet{cui_heres}.

\begin{figure*}
\centering
\includegraphics[width=1.85\columnwidth]{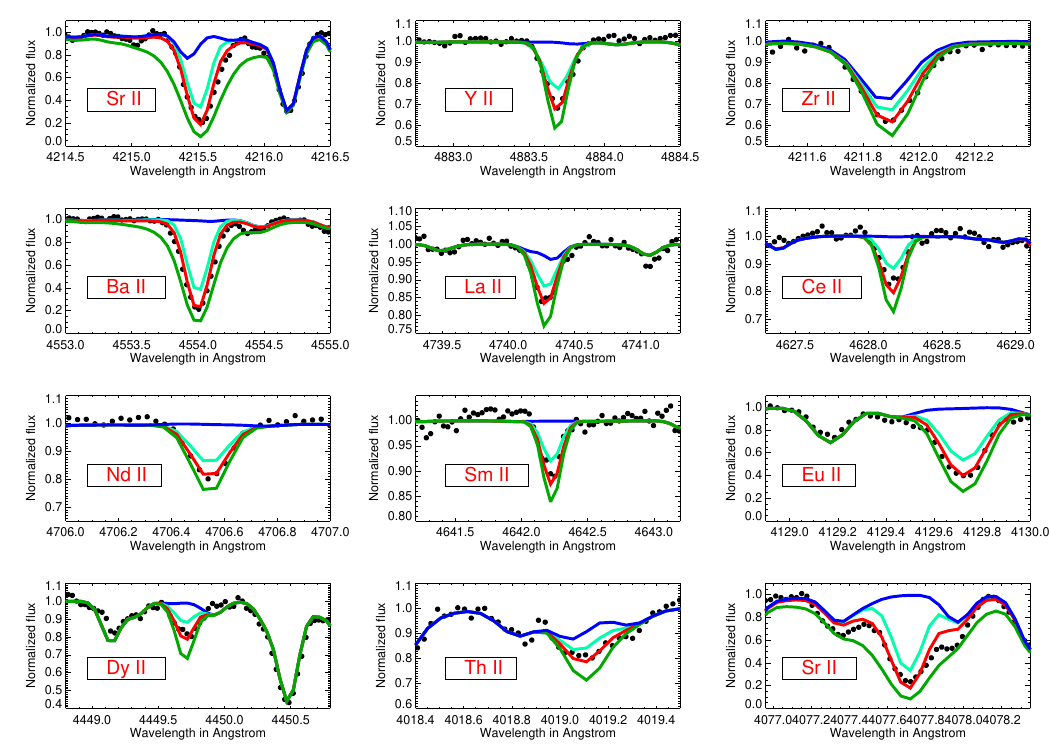}
\caption{Spectral synthesis for the different neutron-capture elements in SDSS J0019+3141. The red line indicates the best fit to the observed points shown in black, while the light blue and green lines correspond to the synthetic spectrum differing by standard uncertainty for the given measurement. The blue line indicates the synthetic spectrum with zero contribution from the given element.}
\end{figure*}

In the case of strontium (Sr), our measurements typically rely on two spectral lines: 
4077\,{\AA} and 4215\,{\AA}. However, the strength of the 4077\,{\AA} line occasionally posed challenges for reliable measurements. Consequently, we utilize the 4215\,{\AA} line exclusively for deriving Sr abundances. For barium (Ba) abundances, we employ the spectral lines at 4554\,{\AA}
and 6141\,{\AA}. Europium (Eu) abundances are successfully derived for SDSS J0019+3141, and an upper limit could be deduced for SDSS J1930+6926 using the 4129\,{\AA} spectral line. Deriving 
thorium (Th) abundances requires special attention, due to the spectral line at 4019\,{\AA} being in proximity to nearby lines of neodymium (Nd) and cobalt (Co). To address this, we determine the abundances of Nd and Co using alternate lines, and held these values constant. We then adjust the Th abundances iteratively to achieve optimal fitting in the spectral region. We could determine Th abundances for only one star, SDSS J0019+3141. Figure 5 provides example syntheses for all the detected $r$-process elements, including Sr, Y, Zr, Ba, La, Ce, Nd, and Th. The fitting of spectral lines in the $r$-II star J0019+3141 is also shown. The red lines are the best-fit values of the corresponding elements, and deviations by $\pm 0.50$ dex are shown with blue and green lines.

\subsection{Uncertainties in Abundances}

The uncertainties linked to the measured abundances are driven by two primary factors: the SNRs of the observed spectra and potential fluctuations in the adopted stellar parameters. To evaluate the influence of SNR, we employ Equation 6 of \citet{cayrel1988} to estimate the uncertainties associated with the abundance determinations. For the stellar parameters, we account for typical uncertainties of approximately $\sim$150\,K in \teff\ and 0.25 dex in $\log g$. These calculated uncertainties are combined in quadrature to yield the errors of the abundance measurements which are the listed in tables 5-15.

\section{Discussion}

The 11 program stars in this study span the metallicity range between [Fe/H] = $-1.65$ and [Fe/H] $ = -3.0$.  Out of the 11 program stars in this study, 8 are VMP stars and one star is an EMP star.  The trends for the abundance ratios in all the detected elements with respect to Fe are shown in Figure 4. The stars from the HESP-GOMPA survey follow the expected trends for all the elements as seen in the figure. The blue-filled circles mark the new stars from this study, while the red-filled star is the $r$-II star SDSS J0019+3141. The abundance patterns for two of the stars (SDSS J0019+3141 and SDSS J0210+3220) are also shown in Figure 6, compared to giants and dwarfs from \citet{cayrel2004} and \citet{cohen2004}, respectively. The average values for the elemental abundances from \citet{cayrel2004} and \citet{cohen2004} are shown in Figure 6 for comparison with the measured abundances. We notice some minor deviations in the expected patterns, but they are within the uncertainties shown by the shaded grey region.

\begin{figure*}
\centering
\includegraphics[width=1.9\columnwidth]{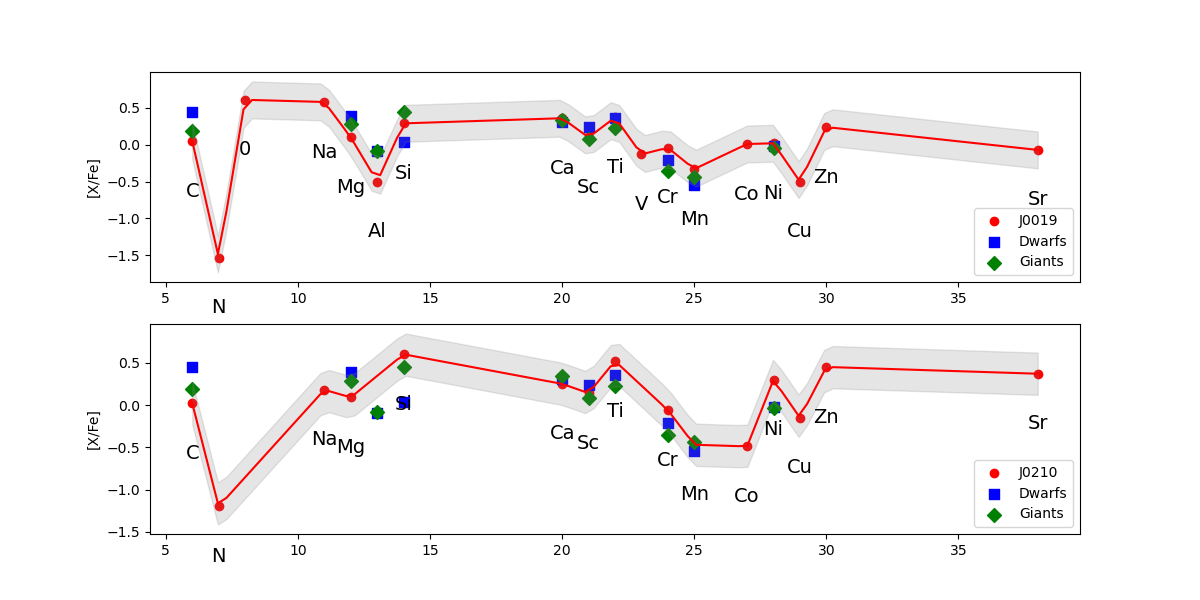}
\caption{The nucleosynthesis pattern for the $r$-II star SDSS J0019+3141 (top panel) and the VMP star SDSS J0210+3220 (bottom panel) are shown in red. The shaded grey region indicates the standard deviation. As a reference, the corresponding mean abundances for giants and dwarf stars from \citet{cayrel2004} and \citet{cohen2004} are also shown with the coloured symbols.}
\end{figure*}

\subsection{$R$-process Origin of Metal-Poor Stars}

Following \citet{tsuji1}, \citet{tsuji2}, and \citet{susmitha}, the light $r$-process elements (e.g., Sr, Y, and Zr) are synthesized by core-collapse supernovae, whereas the heavier $r$-process elements (Ba and beyond) require more energetic phenomena, such as neutron star mergers or similar catastrophic events. The light element primary process (LEPP), more recently known as the limited $r$-process \citep{frebelrev18}, leads to preferential production of the light $r$-process elements in metal-poor stars (see, e.g., 
\citealt{frebelrev18,hansen18rpa,holmbeck18,sakari18rpa,rpa3}). Such a secondary source for $r$-process production is also expected from the existence of the universality and robust nature of the main $r$-process accompanied by significant scatter in the first $r$-process peak. Thus, the light $r$-process/heavy $r$-process ratios indicate the relative contributions of the $r$-process production events and their respective sites. These signatures of $r$-process nucleosynthesis mix with the ISM at the earliest epochs, and are retained in the subsequent metal-poor stellar population, as the contribution from the $s$-process is expected to be minimal in the lowest metallicity regime.  

In this study, we consider Sr as a representative element of the light $r$-process, and Ba as an indicator of the heavier (main) $r$-process. We also use Mg, an $\alpha$-element, as representative of contributions from core-collapse supernovae \citep{lai2008,hegerandwoosley2008}. In figure 7, we demonstrate the [Sr/H] and [Ba/H] abundances as a function of [Mg/H] in black and red respectively, removing the dependency on metallicity [Fe/H]. As seen in the figure, the rise in [Ba/H] with respect to increase in [Mg/H] is sharper than the rise in [Sr/H] as indicated by the difference in their slopes. Hence, as [Mg/H] increases the abundances of Sr and Ba increases at different rates indicating production of the light $r$-process element Sr to be less efficient than heavier $r$-process element Ba with increase in contribution from core collapse supernovae quantified by Mg abundances.

\begin{figure}
\centering
\includegraphics[width=1.0\columnwidth]{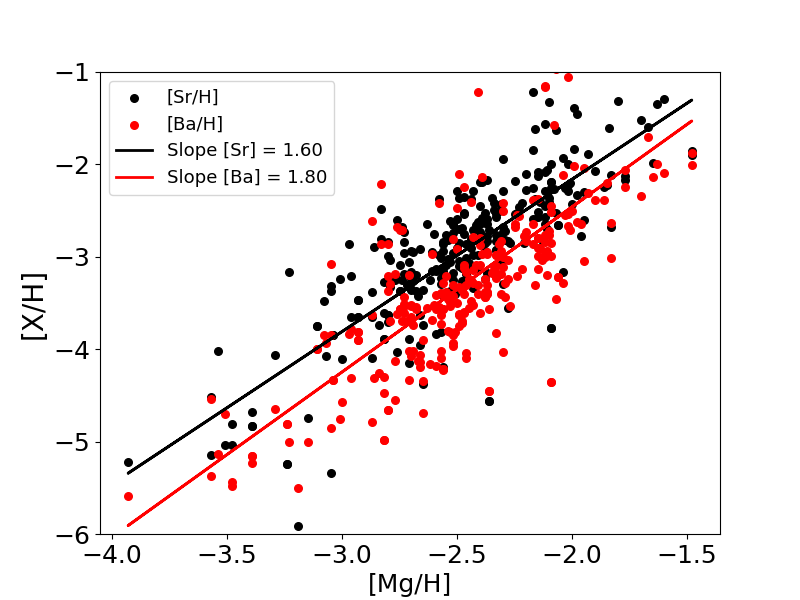}
\caption{The [Sr/H] and [Ba/H] abundances are shown in black and red respectively as a function of [Mg/H] for the metal-poor stars from \citet{roederer2014} and \citet{rpa3}. The slope of their fits are also indicated in the panel.}
\end{figure}

To understand it better, Figure 8 shows the distribution of the [Sr/Ba] abundance ratios as a function of [Mg/H]. We find that, as [Mg/H] increases, [Sr/Ba] initially remains unaltered but then steadily declines as [Mg/H] increases beyond -2.50, as indicated by the red solid line which is the lowess fit to the data taking into account the different fractions of the populations. It is shown to better understand the trends buried within the scatter.The Pearson correlation co-efficient is also negative for all the populations (e.g the value for the $r$-I stars from \citet{rpa3} is $-$0.37 indicating a moderate anticorrelation). The black circles are the non r-process enhanced stars taken from \citet{roederer2014}, while the $r$-I and $r$-II stars taken from \citet{roederer2014} and \citet{rpa3} are shown by blue and purple circles, respectively. The program stars are marked in red. The same downward trend is noticed for all the sub-populations of stars, including the non-RPE stars as seen to be falling within the shaded region, but we also note the large scatter for the $r$-II stars.

The $r$-II stars span a much narrower range than the $r$-I stars in [Mg/H] and appear to be clustered around [Mg/H]=-2.5 beyond which the [Sr/Ba] starts decreasing as shown by the fit. The $r$-process enrichment in $r$-I stars appear to be more gradual with a steadily declining [Sr/Ba] whereas it appears to be more episodic for the $r$-II population. The decreasing trend of [Sr/Ba] with an increase in Mg abundance indicates that the relative contribution of the light to heavy $r$-process decreases as the contribution from core-collapse supernovae increases. The lighter $r$-process element Sr is produced less than the heavier $r$-process element Ba as the contribution from core-collapse supernovae increases. We infer that Type II supernovae may not be the main (or perhaps only) site for preferential production of the light $r$-process elements.

\begin{figure*}
\centering
\includegraphics[width=1.5\columnwidth]{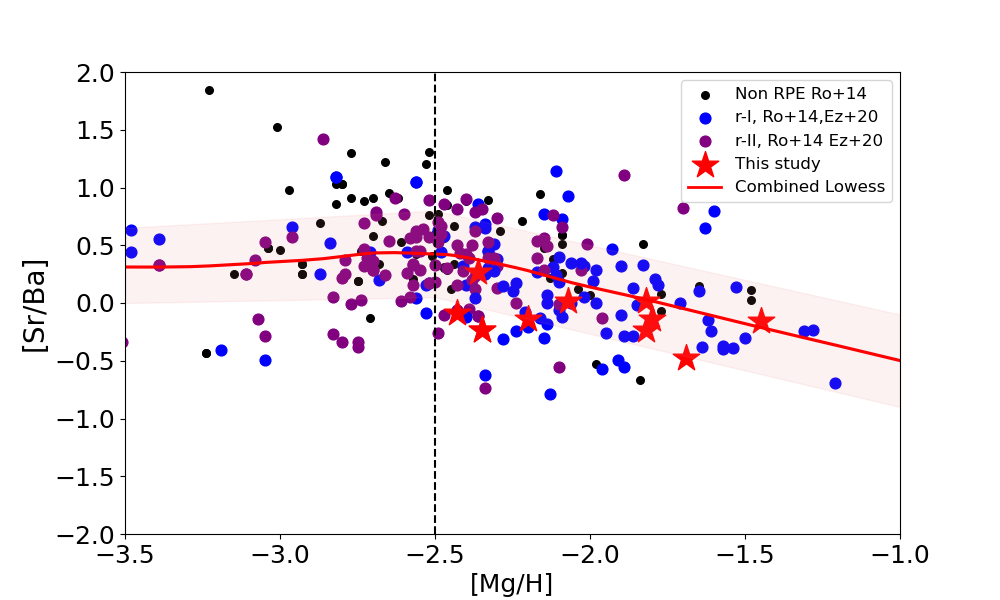}
\caption{The ratio of [Sr/Ba] decreases as [Mg/H] increases (red line), which we interpret as a decrease in the production of light $r$-process with the increasing contribution from Type II supernovae. The red solid line shows the lowess fit to the data. The black dots represent the metal-poor, non RPE halo stars from \citet{roederer2014}, while the $r$-I and $r$-II stars from \citet{roederer2014} and \citet{rpa3} are marked in blue and purple, respectively. The stars in this study are marked with red-filled stars. The black dashed line corresponds to [Mg/H]=-2.5 beyond which the [Sr/Ba] ratio starts declining.}
\end{figure*}

\subsection{The $r$-II Star SDSS J0019+3141}

With an [Eu/Fe] value of +0.78, SDSS J0019+3141 is categorized as an $r$-II star, as per the definition by \citet{rpa4} where $r$-II stars are characterized by [Eu/Fe] $\geq$ +0.70. It stands out as one of the brightest among the known RPE stars. The abundance pattern observed in this star for light elements such as C, $\alpha$-elements like Mg, and the Fe-peak elements (including Fe, Co, and Ni) closely resembles that of typical VMP halo stars. Among the light neutron-capture elements, Sr, Y, and Zr could be measured for this star. The $r$-process abundance pattern, scaled to Solar metallicity and normalized to the Eu abundance, as discussed in \citet{ban_rp}, is shown in Figure 9. The levels and dispersion in the light $r$-process elements suggest the possibility of a limited $r$-process contributing to their formation.

\begin{figure}
\centering
\includegraphics[width=1.05\columnwidth]{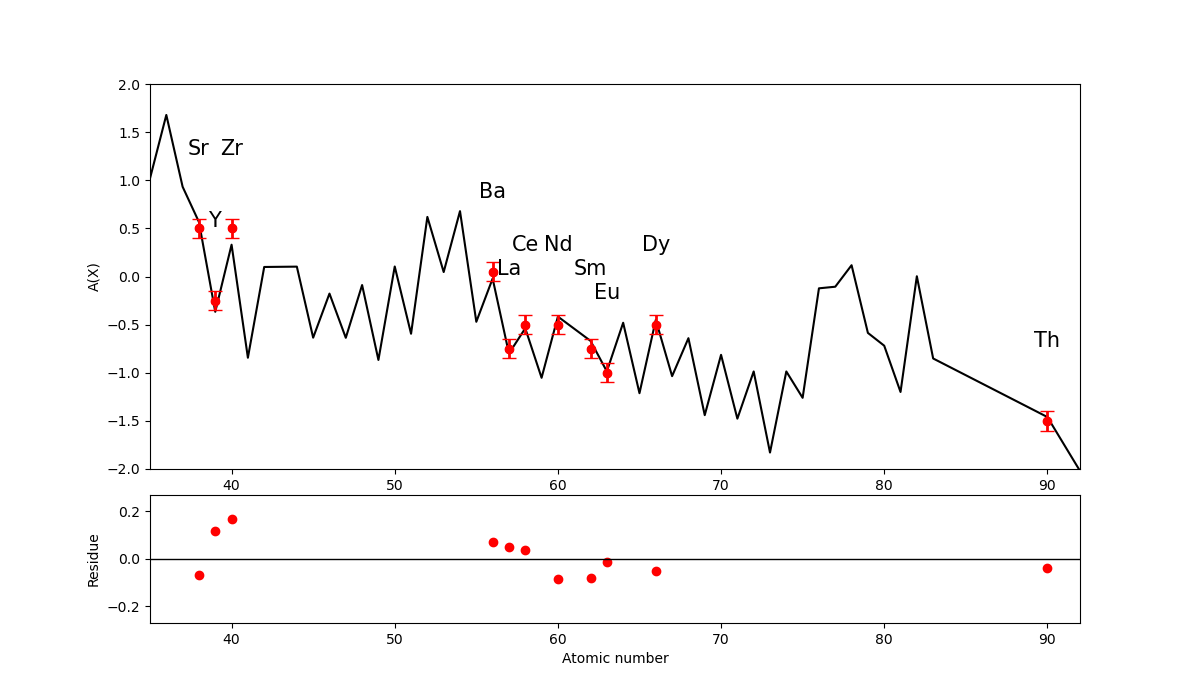}
\caption{Abundance pattern of the neutron-capture elements for SDSS J0019+3141. The scaled-Solar $r$-process abundances are represented by a grey line, while the derived abundances for each specific element, normalized to the abundance of Europium (Eu), are depicted using red-filled circles. The Solar $r$-process abundances used in this analysis were graciously provided by Dr. Erica M Holmbeck (private comm.), based on the computations conducted by \citet{arlandini1999}. The residuals between the abundance measurements and the scaled Solar r-process pattern are shown as a function of atomic number in the lower panel.}
\end{figure}
%<-- NOTE THAT LABELS OF THE ELEMENTS COULD BE LARGER - done

In contrast, the heavy neutron-capture elements in SDSS J0019+3141 exhibit much less scatter and align well with the scaled-Solar abundances, consistent with observations of many other RPE stars in the literature. Thus, the universality and robust nature of the $r$-process pattern for heavy elements is clearly found for this star. With [Ba/Eu] = $-0.61$, the origin of the elements is very likely to be primarily from the $r$-process; the detection of Th is also an indicator of pure $r$-process origin. Regrettably, limitations stemming from the SNR at the blue end of the spectrum restricted our ability to derive abundances for elements approaching the second $r$-process peak. Nevertheless, among the elements for which measurements were possible, including Ba, La, Ce, Nd, Sm, Eu, Dy, and Th, a consistent level of enhancement is evident, in accordance with expectations for the primary $r$-process contribution. Given its relative brightness ($V$ = 10.1), SDSS J0019+3141 is an ideal candidate for higher resolution, higher SNR follow-up spectroscopy to measure the abundances of other key $r$-process elements (Bandyopadhyay et al. (2024)in preparation).

\subsubsection{Cosmochronometry}

The Th abundance in stars is frequently employed as a chronometric tool for the estimation of stellar ages (strictly speaking, the time since the production of the $r$-process elements). The estimation of a star's age can be derived through the application of the relationship \citep{Sneden2002,cain18,Saraf23} 

\begin{equation}
    \Delta t = 46.7[\log \epsilon{\rm (Th/X)}_{i} - \log \epsilon{\rm (Th/X)}_{f}],
    \label{eqn:age_calc}
\end{equation}

\noindent In this equation, where $X$ represents a non-radioactive element, $\log \epsilon {\rm (Th/X)}_{i}$ denotes the initial abundance ratio at the time of the $r$-process event, often referred to as the production ratio (PR), and log $\log \epsilon {\rm (Th/X)}_{f}$ signifies the current abundance ratio. The calculation of the associated error in the age is determined by:

\begin{equation}
    \Delta t_{err} = 46.7\sqrt{\sigma^{2}_{\log \epsilon{\rm (Th)}} + \sigma^{2}_{\log \epsilon{\rm (X)}}},
    \label{eqn:age_calc_err}
\end{equation}

\noindent In this context, $\sigma_{\log \epsilon {\rm (Th)}}$ and $\sigma_{\log \epsilon {\rm (X)}}$ represent the uncertainties associated with the abundance measurements of Th and the stable element X, respectively. Uncertainty also arises from the predicted PR, which depends on theory.

Assessment of $\log \epsilon$(Th/Eu) also serves as an indicator of the actinide-to-lanthanide abundance ratio. In accordance with the categorization of the actinide measurements in stars 
(e.g., \citealt{Holmbeck2019}), J0019+3141 falls within the actinide-normal classification.  It is thus possible to compute a lower limit on its stellar age. The production ratio (PR) values utilized in Equation (1) are taken from \citet{schatz2002} and \citet{hill2017}. Detailed estimations of ages derived from individual elemental species are listed in Table 4. Employing the PR values as reported by \citet{schatz2002} and \citet{hill2017}, the biweight central location of the ages for SDSS J0019+3141 are 13.2 and 17.6 Gyr, respectively. However, we also note a significant difference between the PR values from \citet{schatz2002} and \citet{hill2017} for Th/La and Th/Sm leading to a large discrepancy for the calculated age corresponding to Th/La and Th/Sm.

\begin{table*}
	\caption{Age estimation of SDSS J0019+3141.} 
	\label{tab:Age}
	\begin{tabular}{lccccccr} 
		\hline
		$Th/X$ & J0019+3141 & PR \citep{schatz2002} &Age (Gyr) & PR \citep{hill2017} & Age (Gyr) & $\sigma$ (Gyr)\\
		\hline
		Th/La &$-$0.89 & $-$0.60 &13.5  & $-$0.362 & 24.7 & 11.2 \\
		Th/Ce &$-$1.14 & $-$0.79 &16.3  & $-$0.724 &19.6  & 9.8 \\
		Th/Nd &$-$1.14 & $-$0.91 &10.7 & $-$0.928 & 10.3 & 10.3 \\
		Th/Sm &$-$0.89 & $-$0.61 &13.1  & $-$0.796 & 4.7 & 10.7 \\
		Th/Eu &$-$0.64 & $-$0.33 &14.5  & $-$0.240 & 18.7 & 10.3 \\
		Th/Dy &$-$1.14 & $-$0.89 &11.7  & $-$0.827 &14.5  & 11.2 \\
				\hline 
		Biweight central location & & &13.2 &   &17.6 & \\
		\hline
	\end{tabular}
\end{table*}

\subsection{Chemodynamical Analysis}

\begin{figure*}
\centering
\includegraphics[width=0.9\columnwidth]{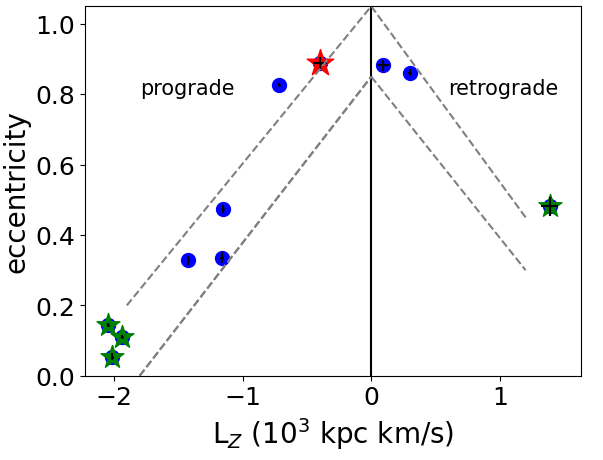}
\includegraphics[width=0.9\columnwidth]{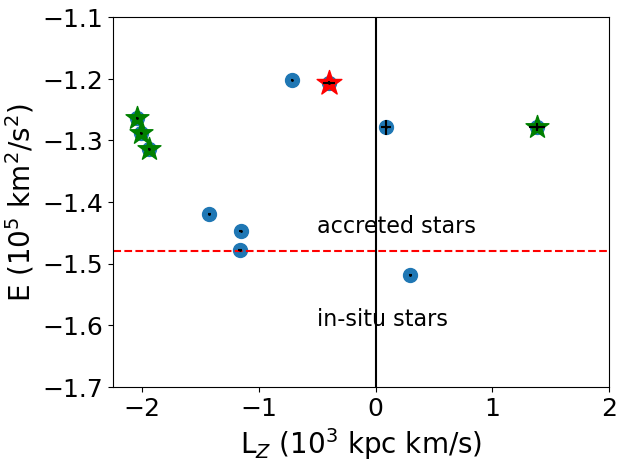}
\caption{Distribution of our program stars in the eccentricity vs. $L_z$ plane (left panel, and in the energy vs $L_z$ plane (right panel). The $r$-II star SDSS J0019+3141 is marked in red, while the $\alpha$-element poor stars are marked in green. The errorbars for each physical quantity for the objects are shown in black. The vertical black line in the left panel indicates the division between prograde and retrograde orbits. The dashed lines in the left panel indicate the region spanned by stars in the study of \citet{nissen} (see text).  The horizontal red-dashed line in the right panel indicates the separation of accreted from in-situ stars.  The vertical black-dashed line in the right panel indicates the division between prograde and retrograde orbits.}
\label{eccen_Lz}
\end{figure*}

\begin{figure*}
\centering
\includegraphics[width=1.86\columnwidth]{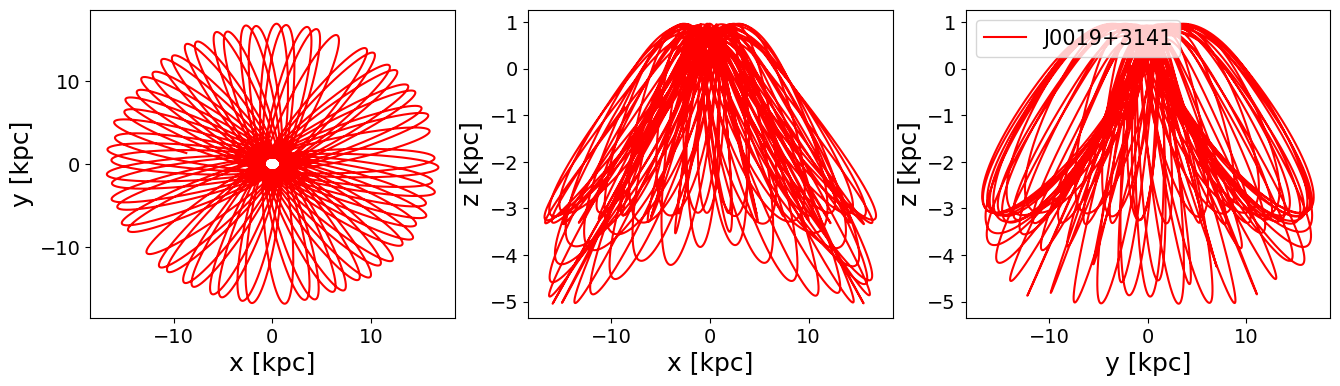}
\includegraphics[width=1.86\columnwidth]{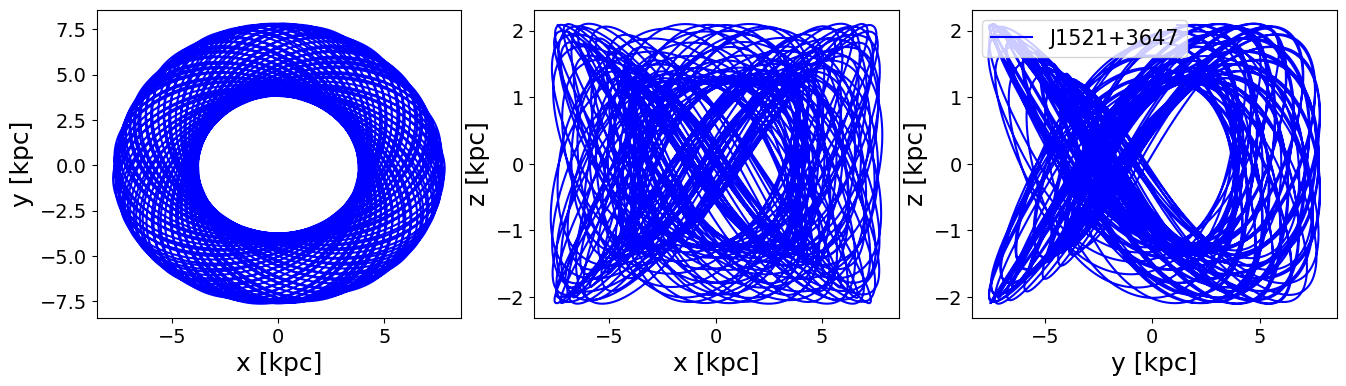}
\includegraphics[width=1.86\columnwidth]{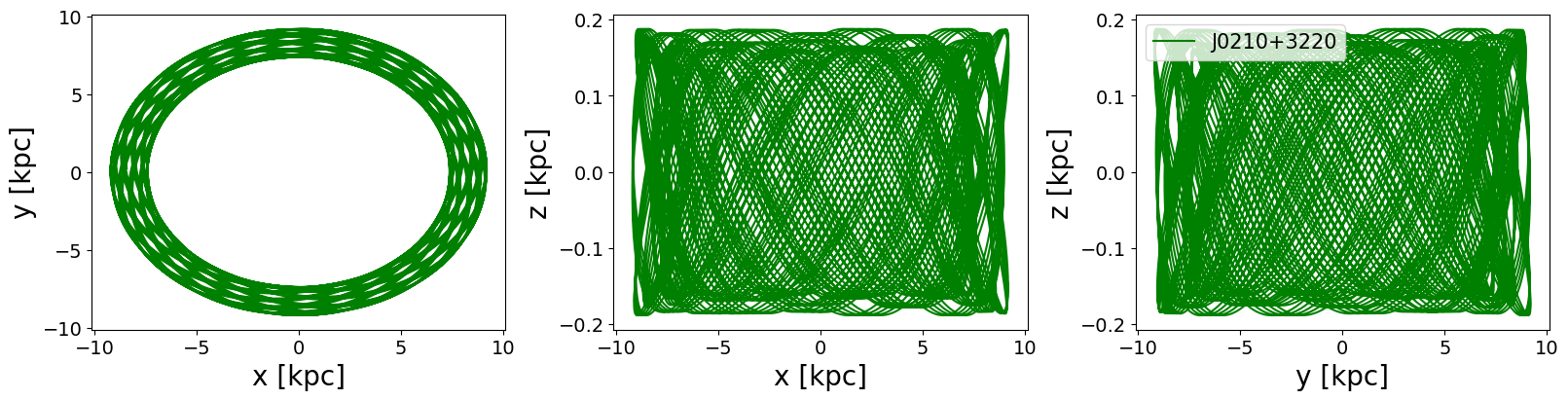}
\caption{Orbital motions of the stars over an orbital integration of 12 Gyr is shown here in the Y-X (left column), Z-X (middle column), and Z-Y planes (right panels). The $r$-II star SDSS J0019+3141 is shown in red in the top panel. Trajectories for the $\alpha$-normal star 
SDSS J1521+3647 (shown in blue) and $\alpha$-poor star SDSS J0210+3220 (shown in green) are in the middle and bottom panels, respectively. }
\end{figure*}

The distribution of the program stars in the eccentricity vs. $L_z$ plane is shown in the left panel of \autoref{eccen_Lz}. %Figure 10. 
The right panel shows their position in the energy vs. $L_z$ plane, which also distinguishes accreted stars from in-situ stars, following \citet{Belokurov23} and other previous studies. We did a Monte Carlo sampling for a sample size of 10000 with multiple iterations for the physical quantities presented in Figure \autoref{eccen_Lz}. The standard deviation of the distribution for the quantities are considered as the uncertainties and shown by black lines on the top of each symbol for the given objects. We also propagated the errors in proper motions, radial velocity, and distances to produce the distributions. The spread in the distributions were found to be small leading to lower uncertainties as seen in the figure. The green-filled stars indicate the $\alpha$-poor stars, red-filled star is the $r$-II star, and the blue-filled dots mark the other stars. The black vertical line in the left panel separates stars with prograde motions from those with retrograde motions. The dotted lines mark the region spanned by the \citet{nissen} stars, the vast majority of which are accreted. Surprisingly, most of the stars in this study also fall in the same space as the \citet{nissen} sample of stars, as seen in the figure. We notice that in the right panel all but one star in this study is most likely to be of accreted origin. Following \citet{haywood2018} and \citet{dimatteo2020}, the low-$\alpha$ stars marked in green are likely to be remnants of an accretion event in the early Galaxy. 

The stars studied here, as well as those in the Gaia DR2 APOGEE sample \citep{dm2019} and the \citet{nissen} sample, are slightly more metal rich than the expected metallicity of the first stars, but as shown by \citet{dimatteo2020}, they can share the same origin despite the differences in metal content \citep{pinto2021}. Most interestingly, the three $\alpha$-poor stars to the left of the figure actually have disc-like orbits, as seen in the Toomre Diagram in Figure 2. Based on the right panel, these stars are also more likely to be accreted. Given that these stars also exhibit prograde motions, this supports the suggestion of  \citet{Carter21} that a reasonable fraction of VMP stars in the disc are accreted from a satellite with the same direction of motion as the disc in the early Galaxy. The $r$-II star also shares a similar accretion history, albeit in the halo.

The trajectories of stars within our sample, classified into in-situ and accreted as delineated in Figure 10, have been calculated using the \texttt{gala} package, which is an Astropy-affiliated tool specifically designed for the study of Galactic dynamics. It is noteworthy that the \texttt{gala} package facilitates the representation of analytical mass models commonly employed in Galactic dynamics, enabling the numerical integration of stellar orbits. These gravitational potential models can be seamlessly integrated with the numerical routines available within the \texttt{gala} package, yielding precise orbital trajectories. Importantly, \texttt{gala} is accompanied by a predefined multi-component model for the Milky Way as described earlier, which proves to be invaluable for orbit computations. For the purposes of our analysis, we have computed these trajectories over a duration of 12 Gyr. 

Figure 11 illustrates the orbital path of the program stars across X, Y, and Z. The $r$-II star is shown in the top panel, while the $\alpha$-poor disc stars are shown in the middle and bottom panels.  It is noteworthy that the trajectory of these particular stars exhibits minimal motion along the Z axis, which can provide important clues for understanding the likely nature of the accreted origin. The $r$-II star also shows motion along the negative Z axis throughout its trajectory, indicating its motion close to the disc but never passing across it. However, we also note that the determination of orbits may not be accurate at the early epochs, and early accretion events may not be properly traced using this technique. 

The distribution of these stars' motion along the $\rho$ (radial distance from the Galactic centre) and Z (vertical distance from the Galactic plane) axes is illustrated in Figure 12. This distribution provides valuable insights into the accretion history of these stars computed for 12 Gyr. Notably, the disc member star shown in the second panel exhibit minimal motion along the Z axis, while the accreted star shown in the third panel span a much larger range along the Z axis. The $r$-II star is shown in the first panel which also shows a significant motion along the Z axis.

\begin{figure*}
\centering
\includegraphics[width=0.65\columnwidth]{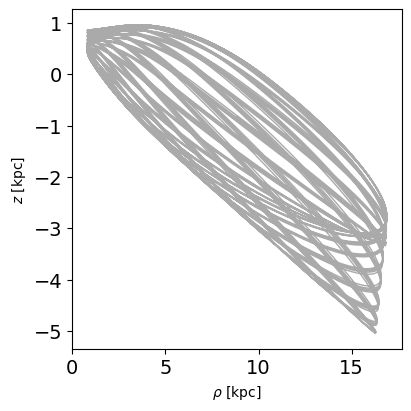}
\includegraphics[width=0.65\columnwidth]{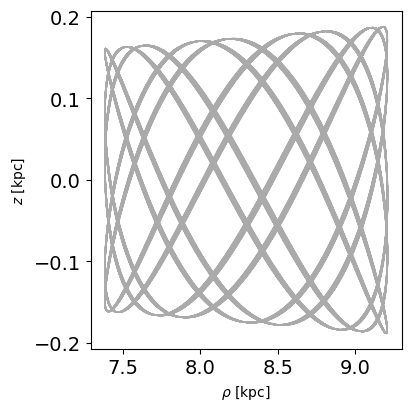}
\includegraphics[width=0.65\columnwidth]{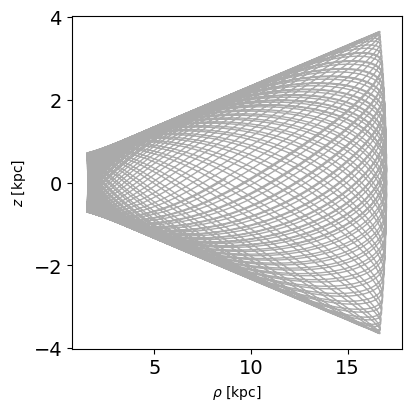}
\caption{The kinematic space occupied by the program stars in the Z vs. $\rho$ plane computed for the same period of 12 Gyr. The representative stars shown in this figure, from left to right, are SDSS J0019+3141, SDSS J0210+3220, and SDSS J0447+5426.}
\end{figure*}

\section{Conclusions}
\label{sec:conclusions}

In the context of the HESP-GOMPA survey, this study reports elemental abundances for 11 metal-poor stars, shedding light on their compositions. These stars exhibit an expected abundance pattern characterised by odd-even nucleosynthesis, consistent with the anticipated characteristics of descendants originating from Type II supernovae. The observed trends across all elements under investigation align with previous research findings, bolstering the credibility of our results. Specifically, the analysis of one star exhibiting an $r$-II signature reveals the universal $r$-process pattern, with notable variations observed in the lighter $r$-process elements. By delving into the origins of $r$-process elements in metal-poor stars, we have demonstrated that the relative production of light $r$-process elements diminishes with an increasing contribution from supernovae to the interstellar medium. Consequently, it appears unlikely that Type II supernovae serve as the sole primary source for these lighter $r$-process elements. Using Gaia astrometric information enabled us to derive the kinematic properties of our sample stars, revealing that all but one are likely of accreted origin. Notably, three of our $\alpha$-poor stars, situated within the halo, exhibit prograde orbits and align with the trajectories of accreted stars from a previous study by \citet{nissen}. This compelling evidence suggests that the disc population contains stars originating from an early accretion event, sharing the same rotational direction as the early MW disc. Additionally, our analysis extends to the derivation of orbits dating back 12 Gyr across distinct planes for the disc and halo, enriching our understanding of their historical dynamics and contributions to the Galactic ecosystem.

% \newpage
\section*{Acknowledgements}

%%%%%%%%%%%%%%%%%%%%%%%%%%%%%%%%%%%%%%%%%%%%%%%%%%
We thank the referee for providing comments that greatly enhanced the quality of the manuscript. We thank the staff of IAO, Hanle, and CREST, Hosakote, who made these observations possible. The facilities at IAO and CREST are operated by the Indian Institute of Astrophysics, Bangalore. T.C.B. acknowledges partial support for this work from grant PHY 14-30152; Physics Frontier Center/JINA Center for the Evolution of the Elements (JINA-CEE), and OISE-1927130: The International Research Network for Nuclear Astrophysics (IReNA), awarded by the US National Science Foundation. R.E. acknowledges support from NSF grant AST-2206263. PKN acknowledges support from the Centro de Astrofisica y Tecnologias Afines (CATA) fellowship via grant Agencia Nacional de Investigacion y Desarrollo (ANID), BASAL FB210003.

\section{Data availability}

The data used in this research will be shared on reasonable request to the corresponding author.

\begin{table*}
	\centering
	\caption{Detailed abundance determinations for SDSS J0019+3141.}
	\label{tab:0019abu}
	\begin{tabular}{lcccccrrrr} % four columns, alignment for each
		\hline
		Element & Species &No. of lines & A(X) & Solar & [X/H] & [X/Fe] & $\sigma$ (dex)\\
		\hline
C$^a$ &CH & \dots   &6.25 &8.43 &$-$2.18 &$+$0.12 &0.18 \\ 
N &CN & \dots &4.0 &7.83 &$-$3.83 &$-$1.53 &0.14\\
O &OI &1 &7.0 &8.69 &$-$1.69 &$+$0.61 &0.17 \\
Na$^b$ &Na I &2 &4.5 &6.24 &$-$1.72 &$+$0.58 &0.11 \\
Mg &Mg I &5 &5.40 &7.60 &$-$2.20 &$+$0.10 &0.16 \\
Al$^b$ &Al I &1 &2.7 &6.45 &$-$3.35 &$-$1.05 &0.19 \\
Si &Si I &2 &5.50 &7.51 &$-$2.01 &$+$0.29 &0.18 \\
Ca &Ca I &11 &4.40 &6.34 &$-$1.94 &$+$0.36 &0.15\\
Sc &Sc II &3 &0.95 &3.15 &$-$2.20 &$+$0.10 &0.18\\
Ti &Ti I &4 &3.1 &4.95 &$-$1.85 &$+$0.45 &0.17 \\
   &Ti II &13 &2.95 &4.95 &$-$2.00 &$+$0.30 &0.13 \\
V &V I &2 &1.50 &3.93 &$-$2.43 &$-$0.13 &0.20 \\
Cr &Cr I &6 &3.10 &5.64 &$-$2.54 &$-$0.24 &0.21 \\
   &Cr II &1 &3.5 &5.64 &$-$2.14 &$+$0.16 &0.22 \\
Mn &Mn I &5 &2.80 &5.43 &$-$2.63 &$-$0.33 &0.14 \\
Co &Co I &2 &2.70 &4.99 &$-$2.29 &$+$0.01 &0.18 \\
Ni &Ni I &4 &3.9 &6.22 &$-$2.32 &$+$0.02 &0.16 \\
Cu &Cu I &2 &1.25 &4.19 &$-$2.94 &$-$0.64 &0.14 \\
Zn &Zn I &1 &2.50 &4.56 &$-$2.06 &$+$0.24 &0.20 \\
Sr &Sr II &2 &0.50 &2.87 &$-$2.37 &$-$0.07 &0.13\\
Y &Y II &2 &$-$0.25 &2.21 &$-$2.46 &$-$0.16 &0.15 \\
Zr &Zr II &3 &0.50 &2.58 &$-$2.08 &$+$0.22 &0.12 \\
Ba &Ba II &2 &0.05 &2.18 &$-$-2.18 &$+$0.17 &0.18 \\
La &La II &3 &$-$0.75 &1.11 &$-$1.86 &$+$0.44 &0.14 \\
Ce &Ce II &3 &$-$0.50 &1.58 &$-$2.08 &$+$0.22 &0.11 \\
Nd &Nd II &2 &$-$0.50 &1.42 &$-$1.92 &$+$0.38 &0.12 \\
Sm &Sm II &2 &$-$0.75 &0.96 &$-$1.71 &$+$0.59 &0.13 \\
Eu &Eu II &1 &$-$1.0 &0.52 &$-$1.52 &$+$0.78 &0.12 \\
Dy &Dy II &2 &$-$0.50 &1.10 &$-$1.60 &$+$0.70 &0.14 \\
Th & Th II & 1 & $-$1.64 &0.06 & $-$1.70 &$+$0.60 & 0.10\\

		\hline
	\end{tabular}
  \newline
    $^a$ Values after applying the corrections due to evolutionary effects from \citet{placco2014}. The C abundance is corrected by 0.75 dex.
    $^b$ Values obtained after applying NLTE corrections. \newline
\end{table*}

\begin{table*}
	\centering
	\caption{Detailed abundance determinations for SDSS J0022+3212.}
	\label{tab:0022abu}
	\begin{tabular}{lcccccrrrr}
		\hline
		Element & Species &No. of lines & A(X) & Solar & [X/H] & [X/Fe] & $\sigma$ (dex)\\
		\hline
C$^a$ &CH & \dots    &6.50 &8.43   &$-$1.93 &$-$0.33 &0.16 \\
Na$^b$ &Na I &2 &4.35 &6.21        &$-$1.86 &$-$0.26 &0.12 \\
Mg &Mg I &3 &6.25 &7.59        &$-$1.34 &$+$0.26 &0.16\\
Ca &Ca I &3 &5.1 &6.32        &$-$1.22 &$+$0.38 &0.16\\
Sc &Sc II &5 &1.53 &3.15       &$-$1.62 &$-$0.02 &0.11\\
Ti &Ti I &7 &3.40 &4.93        &$-$1.53 &$+$0.07 &0.09\\
   &Ti II &6 &3.85 &4.93       &$-$1.08 &$+$0.52 &0.07\\
Cr &Cr I &3 &4.13 &5.62        &$-$1.49 &$+$0.11 &0.08\\
   &Cr II &2 &4.5 &5.62       &$-$1.12 &$+$0.48 &0.06\\
Mn &Mn I &4 &3.4 &5.42        &$-$2.02 &$-$0.42 &0.12\\
Co &Co I &2 &2.9 &4.93        &$-$2.03 &$-$0.43 &0.09\\
Ni &Ni I &3 &4.90 &6.20        &$-$1.30 &$+$0.30 &0.15\\
Cu &Cu I &1 &2.9 &4.19 &$-$1.29 &$+$0.31 &0.17\\
Zn &Zn I &2 &3.30 &4.56        &$-$1.26 &$+$0.34 &0.16\\
Sr &Sr II &2 &1.40 &2.83    &$-$1.43 &$+$0.17 &0.18\\
Ba &Ba II &2 &1.08 &2.25    &$-$1.17 &$+$0.43 &0.14 \\
		\hline
	\end{tabular}
  \newline
    $^a$ Values after applying the corrections due to evolutionary effects from \citet{placco2014}. The C abundance is corrected by 0.01 dex.
    $^b$ Values obtained after applying NLTE corrections. \newline
\end{table*}

\begin{table*}
	\centering
	\caption{Detailed abundance determinations for SDSS J0210+3220.}
	\label{tab:0210abu}
	\begin{tabular}{lcccccrrrr}
		\hline
		Element & Species &No. of lines & A(X) & Solar & [X/H] & [X/Fe] & $\sigma$ (dex)\\
		\hline
C$^a$ &CH & \dots    &6.00 &8.43   &$-$2.43 &$+$0.02 &0.15 \\
Na$^b$ &Na I &2 &3.94 &6.21        &$-$2.27 &$+$0.18 &0.12 \\
Mg &Mg I &4 &5.23 &7.59        &$-$2.36 &$+$0.09 &0.16\\
Si &Si I &1 &6.02 &7.51 &$-$1.49 &$+$0.96  &0.13 \\
Ca &Ca I &8 &4.16 &6.32        &$-$2.16 &$+$0.29 &0.0?\\
Sc &Sc II &5 &0.84 &3.15       &$-$2.31 &$+$0.14 &0.11\\
Ti &Ti I &7 &3.10 &4.93        &$-$1.83 &$+$0.62 &0.09\\
   &Ti II &6 &2.91 &4.93       &$-$2.02 &$+$0.43 &0.07\\
Cr &Cr I &3 &3.06 &5.62        &$-$2.56 &$-$0.11 &0.08\\
   &Cr II &2 &4.15 &5.62       &$-$2.47 &$-$0.02 &0.06\\
Mn &Mn I &4 &1.60 &5.42        &$-$2.92 &$-$0.47 &0.12\\
Co &Co I &2 &1.99 &4.93        &$-$2.94 &$-$0.49 &0.09\\
Ni &Ni I &3 &4.35 &6.20        &$-$1.85 &$+$0.60 &0.16\\
Cu &Cu I &1 &1.59 &4.19 &$-$2.60 &$-$0.15 &0.16\\
Zn &Zn I &2 &2.76 &4.56        &$-$1.80 &$+$0.65 &0.14\\
Sr &Sr II &2 &0.75 &2.83    &$-$2.08 &$+$0.37 &0.17 \\
Ba &Ba II &2 &0.0 &2.25    &$-$2.25 &$+$0.20 &0.15 \\
		\hline
	\end{tabular}
  \newline
    $^a$ Values after applying the corrections due to evolutionary effects from \citet{placco2014}. The C abundance is corrected by 0.0 dex.
    $^b$ Values obtained after applying NLTE corrections. \newline
\end{table*}

\begin{table*}
	\centering
	\caption{Detailed abundance determinations for SDSS J0216+3310.}
	\label{tab:0216abu}
	\begin{tabular}{lcccccrrrr} % four columns, alignment for each
		\hline
		Element & Species &No. of lines & A(X) & Solar & [X/H] & [X/Fe] & $\sigma$ (dex)\\
		\hline
C$^a$ &CH & \dots    &7.0 &8.43   &$-$1.43 &$+$0.22 &0.13 \\
Mg &Mg I &4 &5.90 &7.59        &$-$1.69 &$-$0.04 &0.12\\
Al$^b$ &Al I &1 &2.75 &6.43        &$-$3.68 &$-$2.03 &0.19 \\
Si &Si I &1 &6.4 &7.51 &$-$1.11 &$+$0.54  &0.13 \\
Ca &Ca I &8 &4.8 &6.32        &$-$1.52 &$+$0.13 &0.0?\\
Sc &Sc II &5 &0.88 &3.15       &$-$2.27 &$-$0.62 &0.11\\
Ti &Ti I &7 &3.8 &4.93        &$-$1.13 &$+$0.52 &0.09\\
   &Ti II &6 &3.55 &4.93       &$-$1.38 &$+$0.27 &0.07\\
Cr &Cr I &3 &4.20 &5.62        &$-$1.42 &$+$0.23 &0.08\\
   &Cr II &2 &4.70 &5.62       &$-$0.92 &$+$0.73 &0.06\\
Mn &Mn I &4 &1.77 &5.42        &$-$3.65 &$-$2.00 &0.12\\
Co &Co I &2 &1.66 &4.93        &$-$3.27 &$-$1.62 &0.09\\
Ni &Ni I &3 &4.80 &6.20        &$-$1.40 &$+$0.25 &0.13\\
Cu &Cu I &1 &3.0 &4.19 &$-$1.19 &$+$0.46 &0.18\\
Zn &Zn I &2 &3.35 &4.56        &$-$1.21 &$+$0.44 &0.18\\
Sr &Sr II &2 &0.75 &2.83    &$-$2.08 &$-$0.43 &0.14 \\
Ba &Ba II &2 &0.75 &2.25    &$-$1.50 &$+$0.15 &0.18 \\
		\hline
	\end{tabular}
  \newline
    $^a$ Values after applying the corrections due to evolutionary effects from \citet{placco2014}. The C abundance is corrected by 0.0 dex.
    $^b$ Values obtained after applying NLTE corrections. \newline
\end{table*}

\begin{table*}
	\centering
	\caption{Detailed abundance determinations for SDSS J0447+5426.}
	\label{tab:0447abu}
	\begin{tabular}{lcccccrrrr} % four columns, alignment for each
		\hline
		Element & Species &No. of lines & A(X) & Solar & [X/H] & [X/Fe] & $\sigma$ (dex)\\
		\hline
C$^a$ &CH & \dots    &6.00 &8.43   &$-$2.43 &$+$0.07 &0.19 \\
Na$^b$ &Na I &2 &3.60 &6.21        &$-$2.61 &$-$0.11 &0.12 \\
Mg &Mg I &4 &5.24 &7.59        &$-$2.35 &$+$0.15 &0.16\\
Al$^b$ &Al I &1 &2.80 &6.43        &$-$3.63 &$-$1.13 &0.24 \\
Ca &Ca I &8 &4.04 &6.32        &$-$2.28 &$+$0.22 &0.10\\
Sc &Sc II &5 &0.17 &3.15    &$-$2.98 &$-$0.48 &0.09\\
Ti &Ti I &7 &2.79 &4.93        &$-$2.14 &$+$0.36 &0.12\\
   &Ti II &6 &2.85 &4.93       &$-$2.08 &$+$0.42 &0.07\\
Cr &Cr I &3 &2.90 &5.62        &$-$2.72 &$-$0.22 &0.13\\
   &Cr II &2 &3.30 &5.62       &$-$2.32 &$+$0.18 &0.09\\
Mn &Mn I &4 &2.72 &5.42        &$-$2.70 &$-$0.20 &0.08\\
Co &Co I &2 &2.17 &4.93        &$-$2.76 &$-$0.26 &0.03\\
Ni &Ni I &3 &3.73 &6.20        &$-$2.47 &$+$0.03 &0.05\\
Cu &Cu I &2 &1.71 &4.19        &$-$2.48 &$+$0.02 &0.17 \\
Zn &Zn I &2 &2.10 &4.56        &$-$2.46 &$+$0.04 &0.15\\
Sr &Sr II &2 &0.25 &2.83    &$-$2.58 &$-$0.08 &0.21\\
Ba &Ba II &2 &0.00 &2.25    &$-$2.25 &$+$0.25 &0.19\\

		\hline
	\end{tabular}
  \newline
    $^a$ Values after applying the corrections due to evolutionary effects from \citet{placco2014}. The C abundance is corrected by 0.0 dex.
    $^b$ Values obtained after applying NLTE corrections. \newline
\end{table*}

\begin{table*}
\begin{center}
\caption{Elemental Abundance Determinations for SDSS J0753+4908.} \label{c6t5}
\label{tab:0753abu}
\begin{tabular}{lcccccrrrr}
\hline\hline
Element & Species &No. of lines & A(X) & Solar & [X/H] & [X/Fe] & $\sigma$ (dex)\\
\hline

C$^a$ &CH & \dots    &6.50 &8.43   &$-$1.93 &$+$0.07 &0.16 \\
Na$^b$ &Na I &2 &4.16 &6.21        &$-$2.05 &$-$0.05 &0.18 \\
Mg$^b$ &Mg I &4 &5.77 &7.59        &$-$1.82 &$+$0.18 &0.14\\
Al &Al I &1 &2.28 &6.43        &$-$4.15 &$-$2.15 &0.22 \\
Ca &Ca I &8 &4.30 &6.32        &$-$2.02 &$-$0.02 &0.14\\
Sc &Sc II &5 &1.58 &3.15    &$-$1.57 &$+$0.43 &0.06\\
Ti &Ti I &7 &3.36 &4.93        &$-$1.57 &$+$0.43 &0.08\\
   &Ti II &6 &3.37 &4.93       &$-$1.56 &$+$0.44 &0.07\\
Cr &Cr I &3 &2.93 &5.62        &$-$2.69 &$-$0.69 &0.13\\
   &Cr II &2 &4.64 &5.62       &$-$0.98 &$+$1.02 &0.11\\
Mn &Mn I &4 &3.11 &5.42        &$-$2.31 &$-$0.31 &0.14\\
Co &Co I &2 &2.46 &4.93        &$-$2.47 &$-$0.47 &0.07\\
Ni &Ni I &3 &4.43 &6.20        &$-$1.77 &$+$0.23 &0.04\\
Cu &Cu I &2 &2.65 &4.19 &$-$1.54 &$-$0.46 &0.17\\
Zn &Zn I &2 &3.50 &4.56        &$-$1.06 &$+$0.94 &0.16\\
Sr &Sr II &2 &1.00 &2.83    &$-$1.83 &$+$0.17 &0.21 \\
Ba &Ba II &2 &0.50 &2.25    &$-$1.75 &$+$0.25 &0.20 \\

\hline
\end{tabular}
\end{center}
% \newline
    $^a$ Values after applying the corrections due to evolutionary effects from \citet{placco2014}. The C abundance is corrected by 0.0 dex. \newline
    $^b$ Values obtained after applying NLTE corrections.
\end{table*}

\begin{table*}
	\centering
	\caption{Detailed abundance determination for SDSS J1350+4819.}
    \label{1350abu}
	\label{tab:abundances_TYC1716}
	\begin{tabular}{lccccccccr} % four columns, alignment for each
		\hline
		Element & Species &No. of lines & A(X) & Solar & [X/H] & [X/Fe] & $\sigma$ (dex)\\
		\hline
C$^a$ &CH & \dots   &6.10 &8.43             &$-$2.33 &$+$0.57 &0.14 \\ %0.01 placco
Na$^b$ &Na I &2 &3.75 &6.24                 &$-$2.49 &$+$0.41 &0.16 \\
Mg &Mg I &5 &5.17 &7.60                 &$-$2.43 &$+$0.47 &0.13\\
Ca &Ca I &11 &3.70 &6.34                &$-$2.64 &$+$0.26 &0.05\\
Sc &Sc II &3 &0.25 &3.15                &$-$2.90 &0.0 &0.18\\
Ti &Ti I &4 &2.23 &4.95                 &$-$2.72 &$+$0.18 &0.09\\
   &Ti II &13 &2.00 &4.95               &$-$2.95 &$-$0.05 &0.07\\
Cr &Cr I &6 &2.18 &5.64                 &$-$3.46 &$-$0.56 &0.10\\
Mn &Mn I &5 &2.07 &5.43                 &$-$3.36 &$-$0.46 &0.13\\
Co &Co I &2 &2.37 &4.99                 &$-$2.62 &$+$0.28 &0.07\\
Ni &Ni I &4 &3.45 &6.22                 &$-$2.77 &$+$0.13 &0.09\\
Sr &Sr II &2 &0.00 &2.87               &$-$2.87 &$+$0.03 &0.20\\
Ba &Ba II &2 &$-$0.50 &2.18               &$-$2.68 &$+$0.22 &0.23 \\
		\hline
	\end{tabular}
  \newline
    $^a$ Values after applying the corrections due to evolutionary effects from \citet{placco2014}. The C abundance is corrected by 0.01 dex.
    $^b$ Values obtained after applying NLTE corrections. \newline
\end{table*}

\begin{table*}
	\centering
	\caption{Detailed abundance determination for SDSS J1521+3647.}
	\label{tab:1521abu}
	\begin{tabular}{lccccccccr} % four columns, alignment for each
		\hline
		Element & Species &No. of lines & A(X) & Solar & [X/H] & [X/Fe] & $\sigma$ (dex)\\
		\hline
C$^a$ &CH & \dots    &6.50 &8.43   &$-$1.93 &$+$0.22 &0.15\\
Na$^b$ &Na I &2 &4.19 &6.21        &$-$2.02 &$+$0.13 &0.16 \\
Mg &Mg I &4 &5.77 &7.59        &$-$1.82 &$+$0.33 &0.17\\
Al$^b$ &Al I &1 &2.94 &6.43        &$-$3.49 &$-$1.34 &0.16 \\
Ca &Ca I &8 &4.40 &6.32        &$-$1.92 &$+$0.23 &0.11\\
Ti &Ti I &7 &3.63 &4.93        &$-$-1.30 &$+$0.85 &0.08\\
   &Ti II &6 &3.23 &4.93       &$-$1.70 &$+$0.45 &0.09\\
Cr &Cr I &3 &3.36 &5.62        &$-$2.26 &$-$0.11 &0.07\\
   &Cr II &2 &4.00 &5.62       &$-$1.62 &$+$0.53 &0.06\\
%Mn &Mn I &4 &1.38 &5.42        &$-$4.04 &$-$1.89 &0.02\\
Co &Co I &2 &2.70 &4.93        &$-$2.23 &$-$0.08 &0.08\\
Ni &Ni I &3 &4.06 &6.20        &$-$2.14 &$+$0.01 &0.09\\
Cu &Cu I &2 &2.33 &4.19 &$-$1.86 &$+$0.29 &0.17\\
Zn &Zn I &2 &3.05 &4.56        &$-$1.51 &$+$0.64 &0.16\\
Sr &Sr II &2 &0.50 &2.83    &$-$2.33 &$-$0.18 &0.18 \\
Ba &Ba II &2 &0.25 &2.25    &$-$2.00 &$+$0.15 &0.21\\
		\hline
	\end{tabular}
  \newline
    $^a$ Values after applying the corrections due to evolutionary effects from \citet{placco2014}. The C abundance is corrected by 0.0 dex.
    $^b$ Values obtained after applying NLTE corrections. \newline
\end{table*}

\begin{table*}
	\centering
	\caption{Detailed abundance determination for SDSS J1930+6926.}
	\label{tab:1930abu}
	\begin{tabular}{lccccccccr} % four columns, alignment for each
		\hline
		Element & Species &No. of lines & A(X) & Solar & [X/H] & [X/Fe] & $\sigma$ (dex)\\
		\hline
C$^a$ &CH & \dots   &5.25 &8.43 &$-$3.18 &$-$0.18 &0.14 \\
Na$^b$ &Na I &2 &3.10 &6.21 &$-$2.11 &$-$0.11 &0.10 \\
Mg &Mg I &5 &5.24 &7.59 &$-$2.35 &$+$0.65 &0.12 \\
Al$^b$ &Al I &1 &2.74 &6.43 &$-$3.69 &$-$0.69 &0.14\\
Ca &Ca I &11 &3.88 &6.32 &$-$2.44 &$+$0.56 &0.11\\
Sc &Sc II &3 &$-$0.42 &3.15 &$-$3.57 &$+$0.67 &0.07\\
Ti &Ti I &4 &2.60 &4.93 &$-$2.33 &$+$0.57 &0.10\\
   &Ti II &13 &1.78 &4.93 &$-$3.15 &$+$0.15 &0.08\\
Cr &Cr I &6 &2.59 &5.62 &$-$3.03 &$-$0.03 &0.11\\
   &Cr II &1 &3.50 &5.64 &$-$2.14 &$+$0.86 &0.06\\
Mn &Mn I &5 &2.94 &5.42 &$-$2.48 &$+$0.42 &0.18\\
Co &Co I &2 &1.12 &4.93 &$-$3.81 &$-$0.81 &0.14\\
Ni &Ni I &4 &3.10 &6.20 &$-$3.10 &$-$0.10 &0.12 \\
Sr &Sr II &2 &$-$0.50 &2.83 &$-$3.33 &$-$0.33 &0.17\\
Y &Y II &2 &$-$0.10 &2.21 &$-$2.31 &$+$0.69 &0.16 \\
Zr &Zr II &3 &0.50 &2.59 &$-$2.09 &$+$0.91 &0.19 \\
Ba &Ba II &2 &$-$0.75 &2.25 &$-$3.00 &$+$0.00 &0.22 \\
Ce &Ce II &3 &$-$0.30 &1.58 &$-$1.88 &$+$1.12 &0.24 \\
Eu &Eu II &1 &$-$1.50 &0.52 &$-$2.02 &$+$0.98 &0.20 \\
		\hline
	\end{tabular}
  \newline
    $^a$ Values after applying the corrections due to evolutionary effects from \citet{placco2014}. The C abundance is corrected by 0.01 dex.
    $^b$ Values obtained after applying NLTE corrections. \newline
\end{table*}

\begin{table*}
	\centering
	\caption{Detailed abundance determination for SDSS J1953+4222.}
	\label{tab:1953abu}
	\begin{tabular}{lccccccccr} % four columns, alignment for each
		\hline
		Element & Species &No. of lines & A(X) & Solar & [X/H] & [X/Fe] & $\sigma$ (dex)\\
		\hline
Li &Li I &1 &2.00 & \dots   & \dots   &  \dots  &0.08 \\
C$^a$ &CH & \dots   &6.70 &8.43             &$-$1.73 &$+$0.52 &0.09 \\ 
Na$^b$ &Na I &2 &4.50 &6.24                 &$-$1.74 &$+$0.51 &0.10 \\
Mg &Mg I &5 &5.80 &7.60                 &$-$1.80 &$+$0.45 &0.14\\
Al$^b$ &Al I &1 &3.12 &6.45                  &$-$3.33 &$-$1.08 &0.15 \\
Ca &Ca I &11 &4.60 &6.34                &$-$1.74 &$+$0.51 &0.12\\
Sc &Sc II &3 &1.30 &3.15                &$-$1.85 &$+$0.40 &0.11\\
Ti &Ti I &4 &3.00 &4.95                 &$-$1.95 &$+$0.30 &0.12\\
   &Ti II &13 &2.93 &4.95               &$-$2.02 &$+$0.23 &0.08\\
Cr &Cr I &6 &3.50 &5.64                 &$-$2.14 &$+$0.11 &0.08\\
   &Cr II &1 &3.67 &5.64                &$-$1.97 &$+$0.28 &0.09\\
Mn &Mn I &5 &3.05 &5.43                 &$-$2.38 &$-$0.13 &0.05\\
Co &Co I &2 &2.90 &4.99                 &$-$2.09 &$+$0.16 &0.07\\
Ni &Ni I &4 &4.25 &6.22                 &$-$1.97 &$+$0.28 &0.12\\
Sr &Sr II &2 &1.00 &2.87               &$-$1.87 &$+$0.38 &0.19\\
Zr &Zr II &2 &1.07 &2.58 &$-$1.51 &$+$0.74 &0.16 \\
Ba &Ba II &2 &0.55 &2.18               &$-$1.63 &$+$0.62 &0.13\\
		\hline
	\end{tabular}
  \newline
    $^a$ Values after applying the corrections due to evolutionary effects from \citet{placco2014}. The C abundance is corrected by 0.0 dex.
    $^b$ Values obtained after applying NLTE corrections. \newline
\end{table*}

\begin{table*}
	\centering
	\caption{Detailed abundance determination for SDSS J2320+1742.}
	\label{tab:2320abu}
	\begin{tabular}{lccccccccr} % four columns, alignment for each
		\hline
		Element & Species &No. of lines & A(X) & Solar & [X/H] & [X/Fe] & $\sigma$ (dex)\\
		\hline

C$^a$ &CH & \dots    &6.00 &8.43   &$-$2.43 &$-$0.08 &0.13 \\
Na$^b$ &Na I &2 &3.75 &6.21        &$-$2.46 &$-$0.11 &0.11 \\
Mg &Mg I &4 &5.52 &7.59        &$-$2.07 &$+$0.27 &0.15\\
Al$^b$ &Al I &1 &2.77 &6.43        &$-$3.66 &$-$1.31 &0.24 \\
Ca &Ca I &8 &4.00 &6.32        &$-$2.32 &$+$0.03 &0.13\\
Sc &Sc II &5 &0.06 &3.15    &$-$3.09 &$-$0.74 &0.14\\
Ti &Ti I &7 &3.08 &4.93        &$-$1.85 &$+$0.50 &0.09\\
   &Ti II &6 &2.54 &4.93       &$-$2.39 &$-$0.04 &0.05\\
Cr &Cr I &3 &2.86 &5.62        &$-$2.76 &$-$0.41 &0.09\\
   &Cr II &2 &3.62 &5.62       &$-$2.00 &$+$0.35 &0.07\\
Mn &Mn I &4 &2.54 &5.42        &$-$2.88 &$-$0.53 &0.15\\
Co &Co I &2 &1.88 &4.93        &$-$3.05 &$-$0.70 &0.14\\
Ni &Ni I &3 &4.29 &6.20        &$-$1.91 &$+$0.44 &0.09\\
Zn &Zn I &2 &2.36 &4.56        &$-$2.20 &$+$0.15 &0.08\\
Sr &Sr II &2 &0.50 &2.83    &$-$2.33 &$+$0.02 &0.17 \\
Ba &Ba II &2 &0.00 &2.25    &$-$2.25 &$+$0.10 &0.19 \\
		\hline
	\end{tabular}
 \newline
        $^a$ Values after applying the corrections due to evolutionary effects from \citet{placco2014}. The C abundance is corrected by 0.01 dex.
        $^b$ Values obtained after applying NLTE corrections. \newline
\end{table*}

\clearpage
%\onecolumn

%\bibliographystyle{aasjournal}
%\bibliography{example}

%\newpage
\bibliographystyle{mnras}
\bibliography{mnarx}

%\clearpage
%\twocolumn
%%%%%%%%%%%%%%%%%%%%%%%%%%%%%%%%%%%%%%%%%%%%%%%%%%
% Don't change these lines
\bsp	% typesetting comment
\label{lastpage}
\end{document}